\documentclass[osajnl,showpacs,superscriptaddress,10pt,twocolumn]{revtex4-2}
\usepackage{dcolumn}
\usepackage{bm}
\usepackage{multirow,array}
\usepackage{graphicx}
\usepackage{amsmath}
\usepackage{float}
\usepackage{xcolor}

\begin{document}

\title{High-speed soliton ejection generated from the scattering of bright solitons by modulated reflectionless potential wells}

\author{T. Uthayakumar, L. Al Sakkaf, and U. Al Khawaja}\email{Corresponding author: u.alkhawaja@uaeu.ac.ae}
\affiliation{Physics Department, United Arab Emirates University, P.O. Box 15551, Al-Ain, United Arab Emirates}

\begin{abstract}
We investigate numerically and theoretically the conditions leading to soliton ejection stimulated through the scattering of bright solitons by modulated reflectionless potential wells. Such potential wells allow for the possibility of controlled ejection of solitons with significantly high speeds. At the outset, we describe the scattering setup and characterise the soliton ejection in terms of the different parameters of the system. Then, we formulate a theoretical model revealing the underlying physics of soliton ejection. The model is based on energy and norm exchange between the incident soliton and a stable trapped mode corresponding to an exact solution of the governing nonlinear Schr\"odinger equation. Remarkably, stationary solitons can lead to high-speed soliton ejection where part of the nonlinear interaction energy transforms to translational kinetic energy of the ejected soliton. Our investigation shows that soliton ejection always occurs whenever the incident soliton norm is greater than that of the trapped mode while their energy is almost the same. Once the incident soliton is trapped, the excess in norm turns to an ejected soliton in addition to a small amount of radiation that share translational kinetic energy. We found that higher ejection speeds are obtained with multi-node trapped modes that have larger binding energy. Simultaneous two-soliton ejection has been also induced by two solitons scattering with the potential from both of its sides. An ejection speed almost twice as that of single soliton ejection was obtained. Ejection outcome and ejection speed turn out to be sensitive to the relative phase between the two incoming solitons, which suggests a tool for soliton phase interferometry.
\end{abstract}


\maketitle

\section{Introduction}
\label{introsec}

Solitons are the stable solitary waves that can preserve their shape, amplitude and speed, before and after collisions with barriers, as well as collision with other solitons. This robust nonlinear translation of solitons arises from the precise balance between linear dispersion and nonlinear interaction forces. After its original observation in 1834 on water waves \cite{Russell}, the mathematical description of such exciting phenomenon was formulated during 1895 \cite{Korteweg}. Followed by the pioneering numerical work by Zabusky and Kruskal in 1965, describing the localised stable pulse-like propagation of waves in nonlinear systems, numerous research efforts are made to understand this novel phenomenon \cite{Zabusky}. Many of the naturally occurring waves, to mention a few, ocean waves \cite{Costa}, magnetic domain walls \cite{Victor}, nerve impulse \cite{Heimburg}, tornados and vortices \cite{Desyatniko} belong to the category of solitons. This universal phenomenon with rich dynamics appears in diverse fields of nonlinear science and engineering, namely, ocean waves \cite{Costa}, Bose-Einstein condensates (BEC) \cite{HenrikandChris}, nonlinear photonics \cite{Kivshar}, plasma physics \cite{Kono}, biophysics \cite{Heimburg}, etc.

In general, these solitons are the localised solutions of the nonlinear integrable equations describing nonlinear evolution. Such dynamic evolution of the soliton through the aforementioned nonlinear systems is governed by the nonlinear Schr\"{o}dinger equation (NLSE) \cite{Agrawal,book,Liu}. Notably, in autonomous systems, solitons withhold their initial shape and speed before and after collisions, with a phase shift \cite{Zabusky}. This is apparent due to the absence of time dependence in the nonlinear evolution equations \cite{Tappert, Ponomarenko, Serkin}. However, in real physical situations, soliton interactions are quite complex and experience the external time and space-dependent forces. Furthermore, solitons in these nonautonomous systems can still sustain their initial profile after collisions and adjust to the external potentials, but renouncing the stability in amplitude, speed, and spectra \cite{Chen,Serkin2,Serkin3}. Significant efforts have been made to understand the scattering and interaction dynamics of solitons with varied external potentials, namely, surfaces \cite{surf1,surf2,surf3}, steps \cite{step1,Nogami}, barriers \cite{bar0,bar1,bar5,bar6}, and wells \cite{well1,well2}.

Interaction of bright solitons with attractive potential wells in BEC displayed the dependence of reflection, trapping and transmission on the potential depth. It signified that the overall resonant interaction within the incoming solitons and the bound states of the potential well leads to soliton trapping whereas nonlinear interactions initiate the process of transmission \cite{brand2}. Additionally, the propagation of solitons through a combination of potential wells was exploited to propose a unidirectional flow of solitons \cite{usa1}. Further exploration of cases with PT-symmetric potentials have shed further light on the physical mechanism of such phenomenon resulting from the energy exchange mechanism between the internal modes of the soliton and its centre of mass dynamics during the scattering process \cite{usa2} where trapping was shown to be insignificant  \cite{usa3}. Along this line, discrete soliton-based soliton diode, all-optical switches, logic gates and filters were proposed through suitably modulating the coupling coefficients and adjusting the control soliton power in the waveguide arrays with an effective potential \cite{usa4,usa5}.

However, understanding the physics of translation, ejection and shape formation of the solitons subjected to the diverse external potentials is always a region of special interest. Earlier attempts to understand the interaction of solitons with external potentials reported both the transmission of pulse-like envelopes as well as the breathing at the centre of mass frame. In addition, they revealed that only certain simple potentials allow the preservation of pulse profiles while for other complex potentials the pulse profiles cannot be preserved, resulting in complex shape deformations \cite{Moura}. It is observed for a large BEC trapped in a shallow Gaussian trap with strong confinement in the transverse axis and weakly trapped in the longitudinal axis, the displacement of the condensate is induced by varying scattering length appropriately. When this displaced condensate reaches the edge of the trap, the tail of the condensate can form a single soliton owing to the out-coupling of higher internal energy from BEC cloud and the tendency to eject outward. This process can be repeated by refilling the condensate with the quantity carried away by the ejected soliton. If the quantity of BEC atoms is substantial this would lead to the blowup of solitons from the trap. To understand the resulted features of the soliton emission from the dense BEC trap, an averaged Lagrangian formalism is employed. In the absence of spatial variation in the scattering length, the cloud centre is located at the bottom of the trap corresponding to the fundamental eigenstate. Further, if the critical value related to the threshold for emission of soliton takes more negative values, the deformation of the effective trapping cloud is initiated. This results in oscillation of the atom clouds and the tendency for emission from the trap is commenced. Subsequently, the Gross-Pitaevskii (GP)/NLSE is analysed through a fluid dynamics model that displayed the formation of shock during the early stages of the tunnelling of a BEC droplet bound in a trap. This shock leads to a formation of blip which splits itself from the border of the trap and this ejected portion transforms into a bright soliton. Thereby predicted the possibility of a ``soliton gun" with a particular mass and speed \cite{Dekel,Maria}. An investigation depicting the soliton interaction with an external delta potential leading to the breaking of a soliton into two along with a radiation was also reported, with an estimation for the amplitudes and phases of the transmitted and reflected solitons \cite{Holmer}.

The tunnelling and ejection of solitons through potential barriers has also been demonstrated by launching a Gaussian beam into the trap. For a barrier with a width wider than that of the input beam, there is no transmission through the barrier and the input beam tends to oscillate within the trap. By decreasing the width, tunnelling is initiated linearly through the barrier accompanied by a significant amount of decay in the trapped power with the propagation. The tunnelled beam eventually takes the shape of a nonlocalised wave as it propagates. Experimenting with the Gaussian beam shows that it can transfer some of its power to radiation while shaping a soliton within the trap. The remaining portion of the beam in the trap also behaves as that of a soliton, and intensity of this trapped beam sets the onset of ejection. As long as the initial energy of the beam is lower than that of the barrier energy, the beam remains indefinitely within the barrier. On the other hand, by increasing the initial energy above the energy of the barrier, the beam is allowed to pass through the barrier resulting in soliton ejection \cite{Barak}. Additionally, for a low-power beam introduced into an amplifying trap potential enclosed in a medium with saturable nonlinearity, illustrated linear tunnelling through the trap for weak amplification. When the amplification is sufficiently greater than the tunnelling rate, identical solitons in periodic sequence are ejected from the trap. Finally, for strong amplification, non-periodic multi-soliton ejection is observed \cite{Barak2}. Another interesting study envisaged the manifestation of a soliton-based Newton's cradle in various nonlinear models, where nonlinear absorption stimulates the breaking of higher-order spatial solitons. The best illustration of such a phenomenon is the formation of gap soliton chains in the photonic crystals and Bragg gratings. The specific feature of Newton's cradle involves the formation of a chain of fundamental quasi-solitons under the action of third-order dispersion induced fission on initial N-solitons. During such a process, the tallest soliton generated can travel through the entire chain with consecutive collisions.  This results in the ejection of the soliton with a higher amplitude accompanied by a noticeable frequency shift. Under such inelastic collision, the passing soliton is found to acquire momentum and energy during its translation through the soliton array. The importance of Newton's cradle under the action of the large N-solitons arrays is the power enhancement of the ejected solitons. These ejected solitons can generate a broadband supercontinuum along with the background of multiple dispersive waves and considerable radiations in the background. Further, such system demonstrated a robust dynamics against the action of the stimulated Raman and self-steepening effects as well as dispersive effects greater than the third-order \cite{Katz,Driben}.

Although numerous investigations reported on the mechanism of soliton ejection from external potentials, there are no significant analytical and numerical studies that account for the speed dynamics of the soliton scattered by external potentials. In our study, we investigate bright solitons scattered by modulated reflectionless potential well via the NLSE.  Our preliminary results have revealed, for certain values of the relevant parameters, the ejection of well-defined solitonic pulse featured with constant intensity and width throughout the trajectory. To explain the physics underlying soliton ejection and most importantly to account for the ejection speed, we provide a theoretical model based on energy exchange between the incoming soliton and a trapped mode.  We then shift our investigation towards the understanding of soliton ejection in potential well with multi-node trapped modes. Under this condition, the system is able to eject the solitons with significantly higher speeds compared to the previous situation. We consider also another interesting setup resulting in a simultaneous two-soliton ejection which occurs as a result of two incoming solitons scattered by the potential from both of its sides.

The rest of the paper is organised as follows. In Section II, we present the characterisation of the soliton ejection in terms of the parameters of the input soliton and the potential. In Section III, we formulate the theoretical model that explains the physics of the ejection mechanism and accounts for the ejection speed. The soliton ejection mechanism of the multi-node trapped modes and the two-soliton ejection are investigated in Section IV. Lastly, Section V provides the conclusions and discussions of our findings in the proposed study.

\section{Characterisation of soliton ejection}
\label{charasec}
Here we describe the set up and parameter regime that lead to soliton ejection. We also investigate the dependence of the ejection speed on the various parameters of the system.

Typically, when a bright soliton of the NLSE is scattered by a reflectionless potential well, such as the P\"oschl-Teller potential, full transmission or full reflection takes place \cite{goodman}. The two scattering outcomes are separated by a sharp transition taking place at a specific critical speed below which, full (quantum) reflection occurs and above which the soliton fully transmits the potential well \cite{brand2,brand1}. At the critical speed, an unstable trapped mode is formed where the energy and norm of the incident soliton are equal to those of the trapped mode \cite{usa6}. In both full transmission and full reflection cases, a small trapped mode is formed temporarily during the scattering process which is then evacuated out of the potential to join the scattered soliton.  For such a case, no soliton ejection occurs and the speed of the scattered soliton, whether reflected or transmitted, is equal to the speed of the incident soliton. In contrast, soliton ejection occurs when part of the incident soliton forms a stable trapped mode while the remaining part is ejected from the potential mainly in the form of a soliton accompanied by a small amount of radiation. The trapped mode turns out to be a stable stationary state of the NLSE in the presence of the reflectionless potential. It has a lower energy and norm compared with the incident soliton. In a sense, the effect can be looked at as a transition from an energy level corresponding to the incident soliton to a lower energy level corresponding to the trapped soliton where the difference is converted to translational kinetic energy of an ejected soliton. It is interesting to see that the difference in energy which is essentially in the form of nonlinear interaction is converted to translational kinetic energy. Our numerical simulations have indeed shown that, for soliton ejection to take place, the norm of the incident soliton should be larger than the norm of the trapped mode. We found also that the potential has to deviate from being reflectionless for this effect to be realised. Specifically, the width of the potential has to be larger than what a reflectionless potential should have, as we will detail shortly.

\begin{figure}[bt]\centering
	\includegraphics[width=5cm,clip]{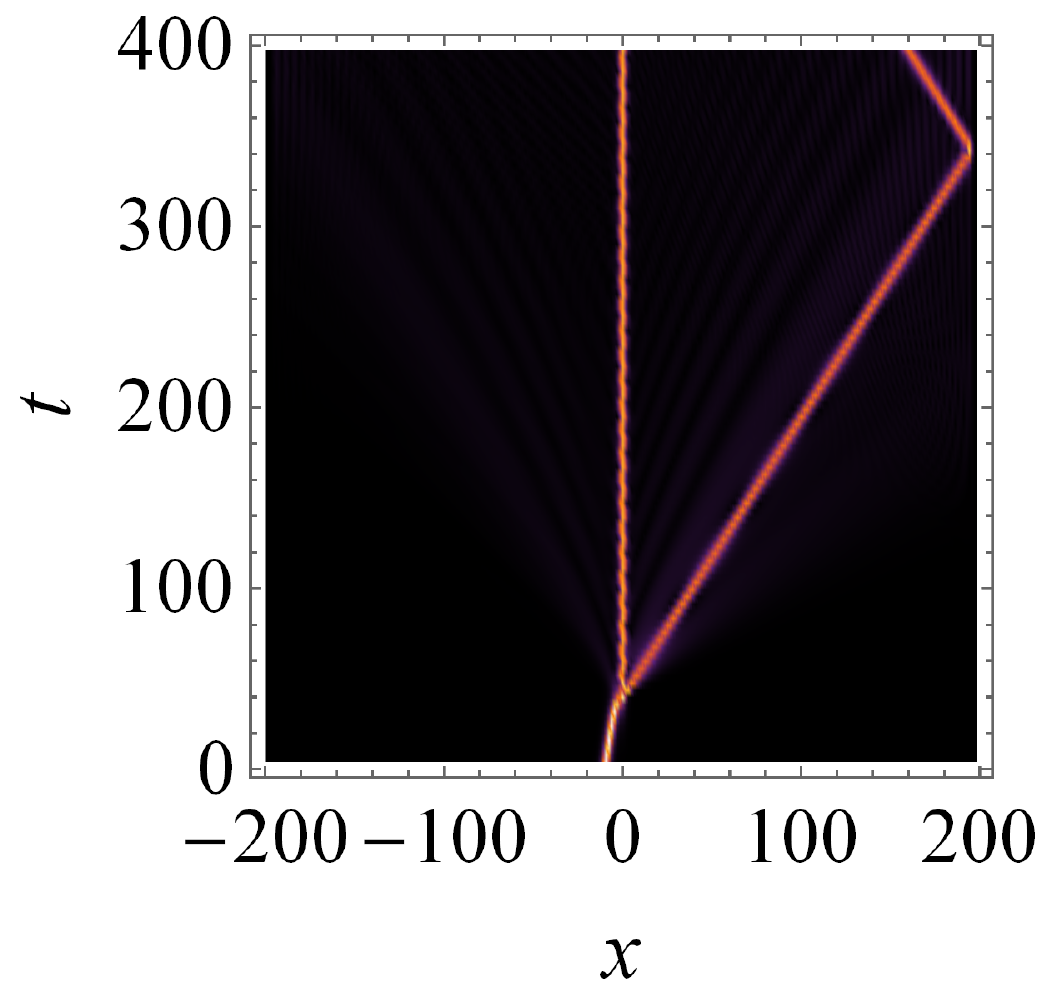}
 \caption{Spatio-temporal plot showing soliton ejection from the potential well with $V_0=3.9$ and $\alpha=0.69\sqrt{V_0}$ with an ejection speed $v_e=0.32$. Other parameters: $u_0$ = 1, $x_0 = -10$, $v_i=0.15$, $g$ = 1. }
  \label{fig1}
\end{figure}

Soliton scattering dynamics is governed by the NLSE \cite{HenrikandChris},
\begin{eqnarray}\nonumber
i\frac{\partial}{\partial t}\psi(x,t)&=&-\frac{1}{2}\frac{\partial^2}{\partial x^2}\psi(x,t)-g|\psi(x,t)|^2\,\psi(x,t)\\&&+V(x)\,\psi(x,t)
\label{nlse},
\end{eqnarray}
where $\psi(x,t)$ is the field describing the intensity of the soliton. For matter-wave solitons in Bose-Einstein condensates, it corresponds to the wave function of the condensate. Here, $g$, is a positive constant determining the strength of the cubic nonlinearity.
The potential well is the P\"oschl-Teller reflectionless potential
\begin{equation}
V(x)=-V_0\,{\rm sech}^2(\alpha\,x)
\label{pot},
\end{equation}
where the depth, $V_0$, of the potential and its inverse width, $\alpha$, are related by $V_0=\alpha^2$. However, as we mentioned above, this reflectionless potential condition is to be violated in order to obtain soliton ejection, namely we will use $\alpha\ne\sqrt{V_0}$.
The incident soliton is taken as the exact bright soliton solution of the fundamental NLSE, i.e. Eq. (\ref{nlse}) without the potential
\begin{eqnarray}\nonumber
\psi(x,t)&=&{\frac{u_0}{\sqrt g}}\,\,{\rm sech}[u_0(x-x_0-v_i\,t)]\\ &&\times e^{i[v_i(x-x_0)+\frac{u_0^2-v_i^2}{2}(t-t_0)]}
\label{psiint},
\end{eqnarray}
where $x_0$, $u_0$, and $v_i$ are arbitrary real constants representing the soliton initial position, amplitude, and speed. It is assumed that the soliton is initially located at the left side of the potential, $x_0<0$, and is launched towards the potential with initial speed $v_i>0$. The scattering outcome is determined by solving numerically Eq. (\ref{nlse}) with $\psi(x,0)$ from Eq. (\ref{psiint}) as an initial profile. The scattered intensities are quantified by calculating the scattering coefficients

\begin{equation}
R=(1/N)\int_{-\infty}^{-l}|\psi(x,t)|^2\,dx,
\end{equation}
\begin{equation}
T=(1/N)\int_{l}^{\infty}|\psi(x,t)|^2\,dx,
\end{equation}
\begin{equation}
L=(1/N)\int_{-l}^{l}|\psi(x,t)|^2\,dx,
\end{equation}
where $R$, $T$, and $L$ are the scattering coefficients of reflection, transmission, and trapping, respectively. Here, $2l$ encompasses the potential region which can be guaranteed by taking for instance $l\approx\,5/\alpha$. Normalisation of the soliton is defined by $N=\int_{-\infty}^{\infty}|\psi(x,t)|^2\,dx$.

In Fig. \ref{fig1}, we show a typical soliton ejection case. It is clear that the well-localised ejected pulse is solitonic since its intensity and width are preserved along its trajectory. It is also noticed that an amount of radiation is emitted as well, which is due to breaking the reflectionless criterion of the potential. Interestingly, the ejection speed is significantly larger than the speed of the incident soliton. In fact, we will show below that high-speed ejection is possible even with stationary input solitons.

\begin{figure}[bt]\centering
\includegraphics[width=8cm,clip]{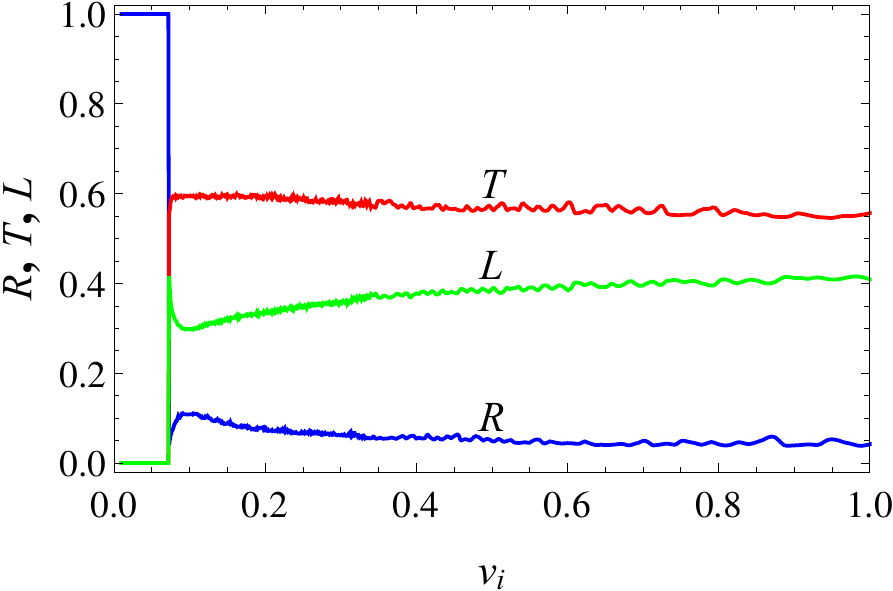}
 \caption{Scattering coefficients of soliton with initial amplitude $u_0$ = 6.3 using the considered potential well in terms of incidence speed. Other parameters: $g = 1$, $V_0 = 100$, $x_0 = -15$, $\alpha = 0.69\sqrt{V_0}$. }
  \label{fig3}
\end{figure}

It is instructive to know the size of the ejected soliton. Therefore, we calculate the scattering coefficients long after scattering which show in Fig. \ref{fig3} that the norm of the ejected soliton is about half of the incident norm ($\approx56\%$), while the other half is mainly trapped ($\approx40\%$), and a small portion is emitted in the form of radiation ($\approx4\%$). Similar to the ejection speed, the norm of ejected soliton is almost insensitive to the speed of incidence for incidence speeds away from the critical speed for quantum reflection at $v_i\approx0.075$.

In the following, we present a detailed investigation of the parameters that affect the value of the ejection speed. This will help to identify regimes in the parameters space where high-speed ejection can be obtained. Specifically, we will investigate the effects of initial soliton speed $v_i$, initial soliton position $x_0$, its initial amplitude $u_0$, and the potential depth $V_0$. \\

\subsection{Effect of the initial speed of input soliton}
\begin{figure*}[bt]\centering
\includegraphics[width=12cm,clip]{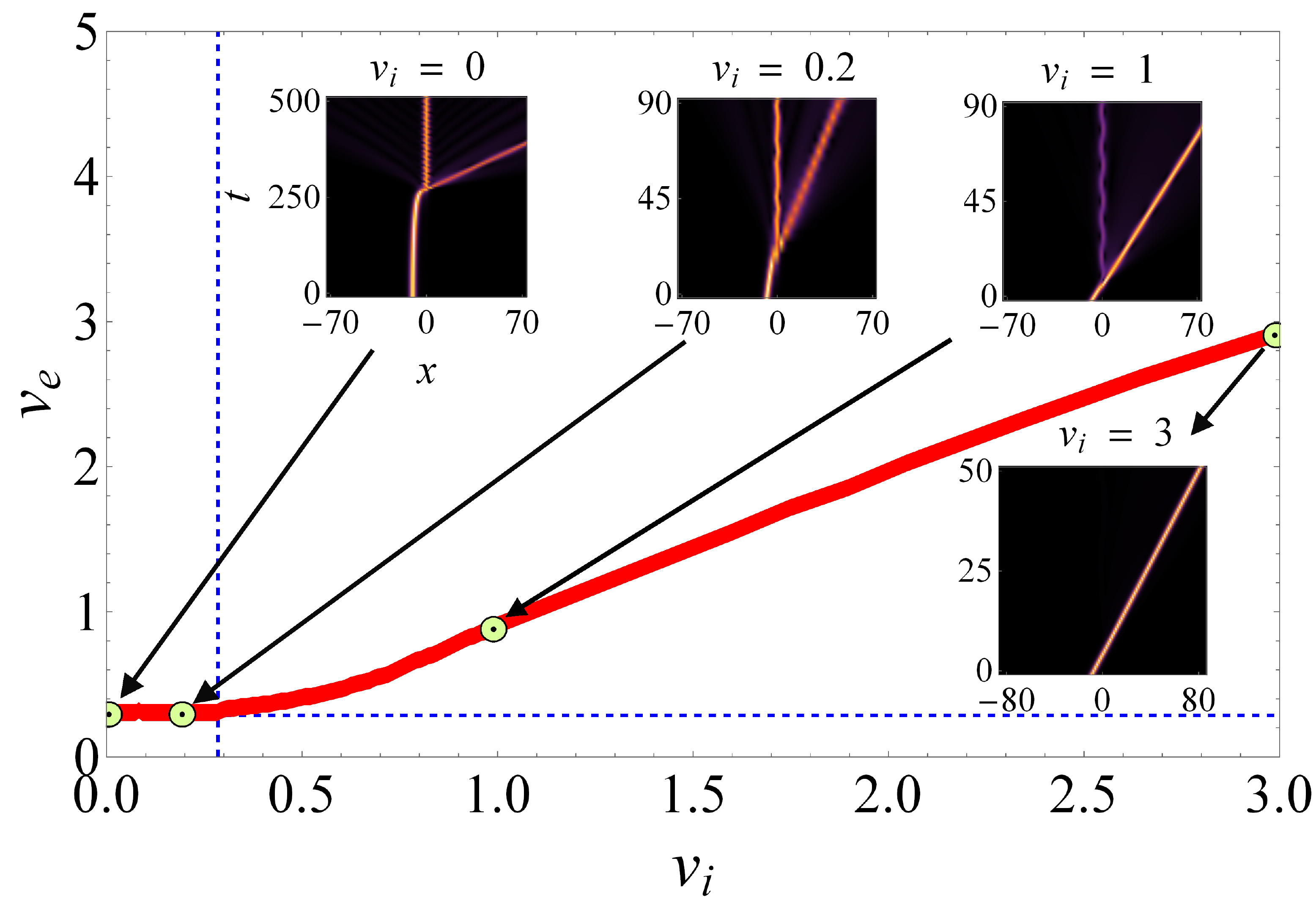}
 \caption{Speed of the ejected soliton propagated with different initial velocities. Spatio-temporal inset figures  describe the dynamics of the soliton propagated with certain specific initial velocities. Constant ejection speed $v_e=0.32$ is obtained for incident velocities up to the vertical dotted line at incident soliton speed $v_i=0.285$. Parameters: $x_0 = -10$, $u_0 = g = 1$, $V_0 = 2$, $\alpha = 0.69\sqrt{V_0}$.}  \label{fig2}
\end{figure*}

Here, we investigate the effect of input soliton's initial speed on the appearance or non-appearance of soliton ejection as well as on ejection speed. We consider a fixed initial position and vary the speed over a range that comprises all different possible outcomes. In Fig. \ref{fig2}, we plot the speed of the ejected soliton, $v_e$, versus the speed of incidence, $v_i$. The figure shows that $v_e$  acquires a constant value for $0<v_i<0.285$ followed by an almost linear increase for large $v_i$. The spatio-temporal inset figures show that soliton ejection is gradually disappearing for large $v_e$. It is appearing for the smaller values of $v_e$ where the speed gain $v_e/v_i$ is significant.

\subsection{Effect of the initial position of a stationary input soliton}
\begin{figure*}[t]\centering
\includegraphics[width=12cm,clip]{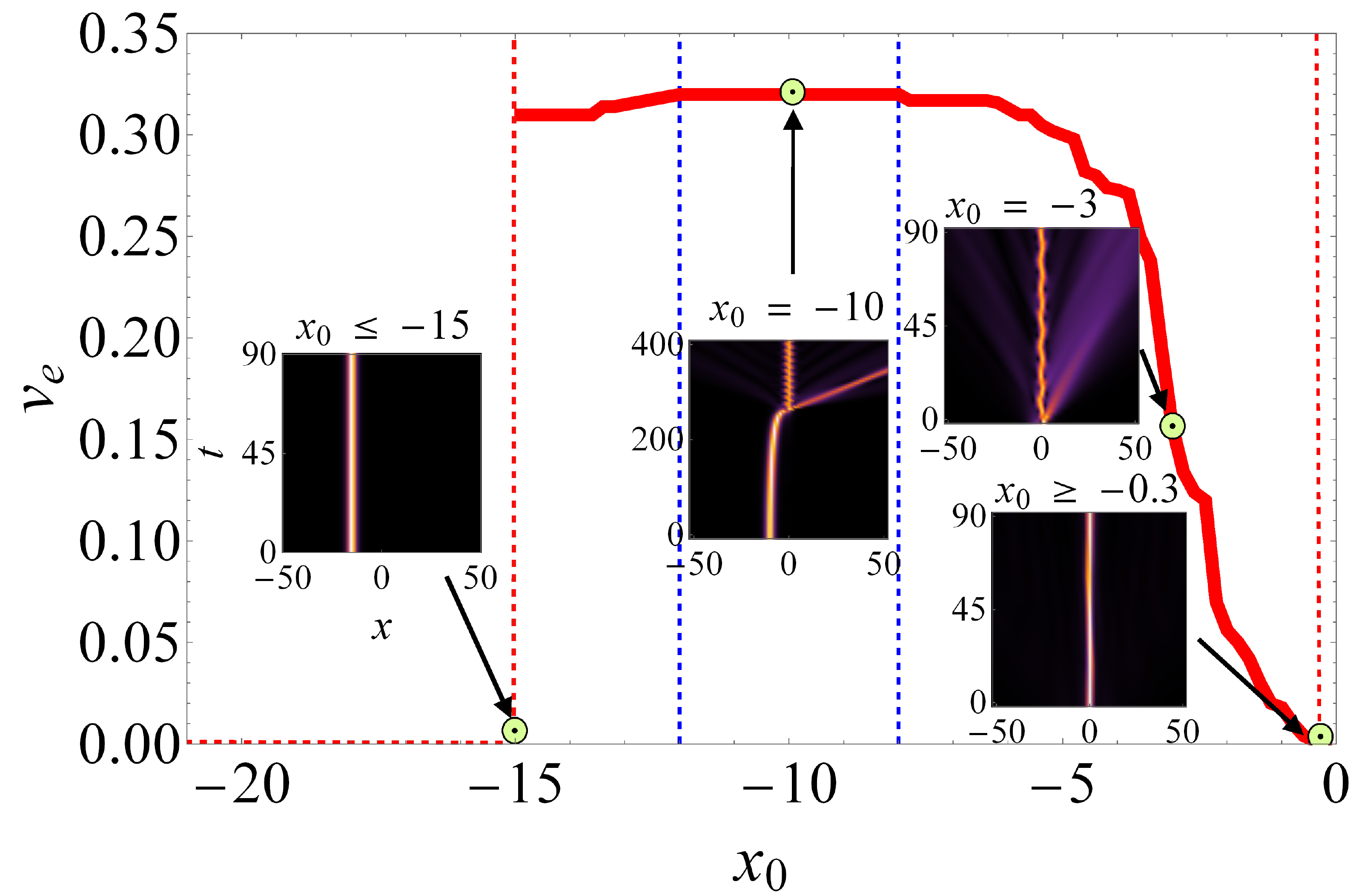}
 \caption{Speed of ejected soliton versus the position of stationary input soliton. Spatio-temporal inset figures describe the dynamics of the soliton propagated from certain specific initial positions. Vertical dotted lines show the boarders at which soliton ejection disappears ($x_0<-15$ and $x_0>-0.3$), and the boarders within which the ejection speed is constant ($-12.1<x_0<-8$). Parameters: $u_0 = g = 1$, $V_0 = 2$, $\alpha = 0.69\sqrt{V_0}$.}
  \label{fig4}
\end{figure*}

Interestingly, soliton ejection can be obtained even with a stationary input soliton. However, if the input soliton is too far from, or too close to the potential well, soliton ejection does not occur. Here, we consider initial positions ranging from $x_0=-20$ to the centre of the potential well and using the specific set of parameters $u_0 = 1$, $g = 1$, $V_0 = 2$ and $\alpha = 0.69\sqrt{V_0}$. The results obtained are provided in Fig. \ref{fig4}. For $x_0\leq-15$, the input soliton is not affected by the potential and thus remains stationary at its initial position. This is indicated by the region before the dashed vertical red line in Fig. \ref{fig4} and the inset plot portrays its dynamics at this situation. For $x_0 > -15$, soliton ejection starts to take place with ejection speed $v_e=0.31$. The spatio-temporal plot in the inset for $x_0= -10$ clearly shows the soliton ejection from the potential well.  A constant ejection speed of $v_e=0.32$ is observed for a rather wide range of $x_0$, varying from $-12.1$ to $-8.0$, which is indicated by the region in between the vertical blue lines of Fig. \ref{fig4}. For initial positions closer to the potential, $v_e$ decreases to reach a minimum of $0.0015$ at $x_0 = -0.4$ with significantly high radiation loss. This portrays the highly attractive dynamics of the potential well which results in the increased soliton trapping. The complete trapping of the soliton is observed for the stationary soliton positioned at $x_0 = -0.3$, as shown in the inset plot.

\subsection{Effect of the initial position of a moving input soliton}
\begin{figure*}[t]\centering
\includegraphics[width=12cm,clip]{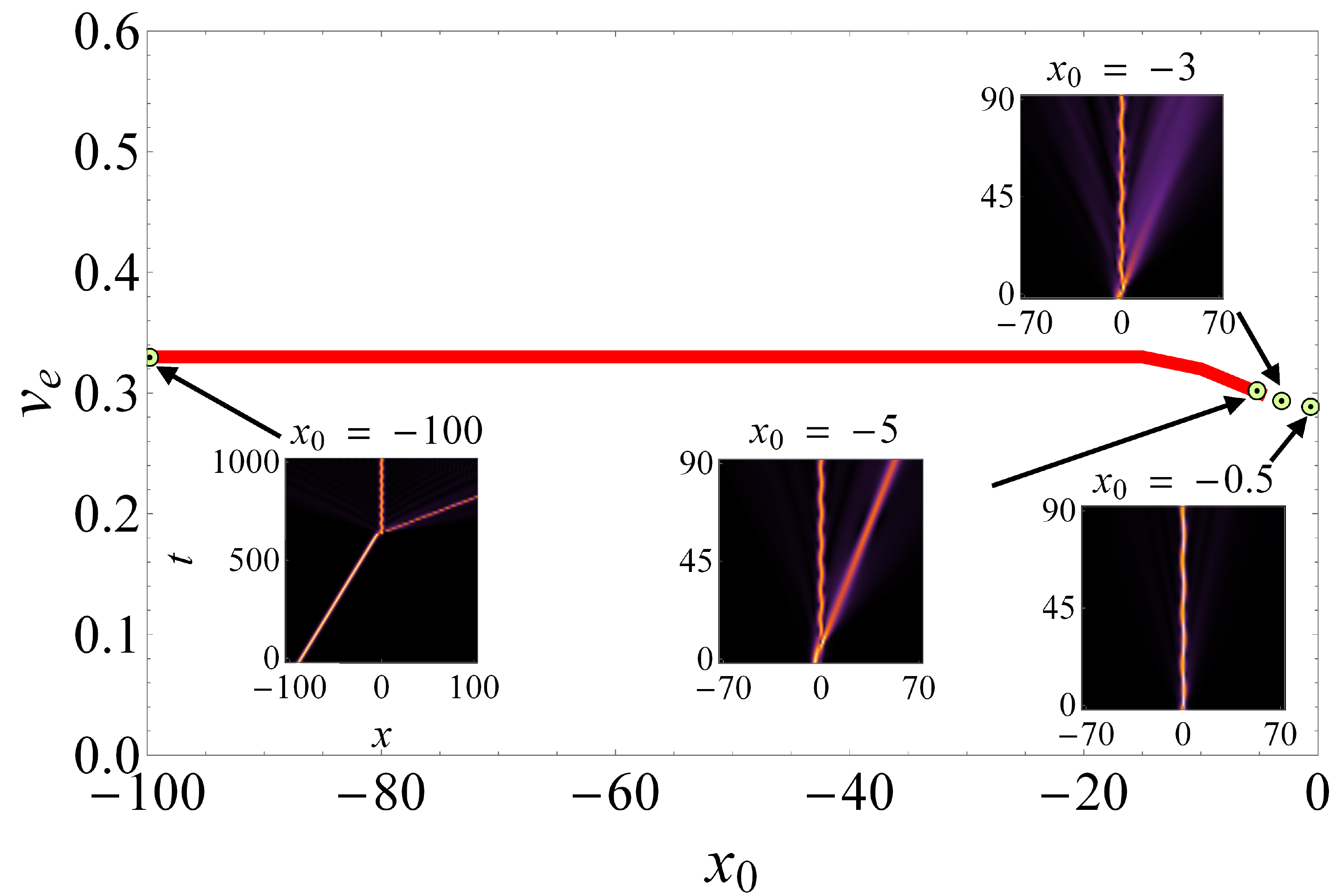}
 \caption{Speed of the ejected soliton for soliton propagating from different initial positions with $v_i= 0.15$. Spatio-temporal inset figures describe the dynamics of the soliton propagated from certain specific initial positions. Other parameters: $u_0 = g = 1$, $V_0 = 2$, $\alpha = 0.69\sqrt{V_0}$.}
  \label{fig5}
\end{figure*}

The aim here is to investigate the effect of input soliton speed on the results of the previous case. Obviously, with a finite speed, the input soliton will reach the potential after some time. Thus, the nonappearance of soliton ejection for large initial positions will not occur here; ejection will eventually always take place, as long as the input soliton is not too close to the potential.  We consider the same set of parameter values as in the previous case with an initial soliton speed $v_i$ = 0.15. The initial position of the soliton is varied from $x_0=-100$ to the centre of the potential well. From the dynamics, as displayed in Fig. \ref{fig5}, solitons are found to eject with a constant speed over a wide range of initial positions ranging from $x_0= -100$ to $x_0=-15$. Throughout this range of initial positions, the speed of ejection is $v_e=0.33$, with a constant speed gain of $v_e/v_i=2.2$. Thereafter, $v_e$ reduces to reach a minimum value of $v_e=0.3$ for $x_0 = -5$. For initial positions closer to the potential, $x_0>-5$, soliton ejection mechanism is lost and trapping with a small amount of radiation dominates. For $x_0\geq-0.5$, the soliton is completely trapped in the potential well, as shown in the inset plot.

\subsection{Effect of input soliton amplitude}
\begin{figure*}[t]\centering
\includegraphics[width=12cm,clip]{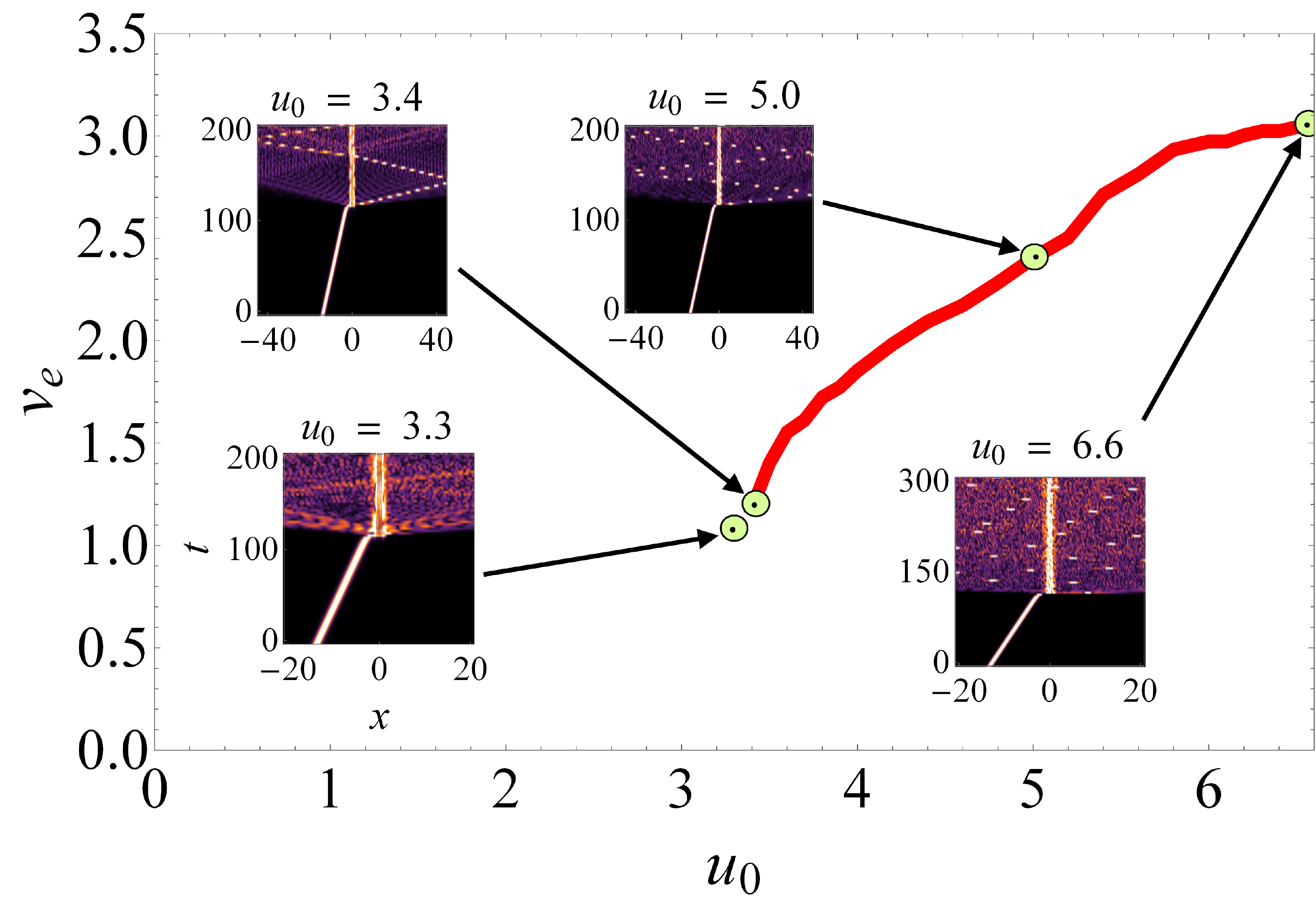}
 \caption{Speed of the ejected soliton for different input amplitudes. Spatio-temporal inset figures describe the dynamics of the soliton propagated with certain specific initial amplitudes. Trajectory of ejected soliton appears dotted due to the high soliton speed which misses some snapshot frames. Parameters: $x_0 = -15$, $v_i = 0.1$, $g = 1$, $V_0 = 100$,  $\alpha = 3.5\sqrt{V_0}$. }
  \label{fig6}
\end{figure*}

Soliton ejection dynamics for input soliton with different initial amplitudes is analysed in this section. For the study, we use the parameters $x_0 = -15$, $v_i = 0.1$, $V_0 = 100$, and $\alpha=3.5\sqrt{V_0}$. For input soliton amplitudes up to $u_0 = 3.3$, no clear soliton ejection takes place since soliton trapping in the potential well is found to be dominant, which is also accompanied with a considerable amount of radiation, as shown in Fig. \ref{fig6}. Soliton ejection is initiated around $u_0=3.4$ with an ejection speed $v_e = 1.18$ and gain $v_e/v_i=11.8$. The soliton ejection dynamics at this initial amplitude is provided in the inset plot. For further increase in $u_0$, the ejection speed is found to increase considerably. A clear soliton ejection regime with low radiation is obtained up to $u_0 = 6.6$ with a speed gain exceeding $30$. Thereafter, an increased amount of radiation dominates the scattering outcome.

\subsection{Effect of potential depth}
\begin{figure*}[t]\centering
\includegraphics[width=12cm,clip]{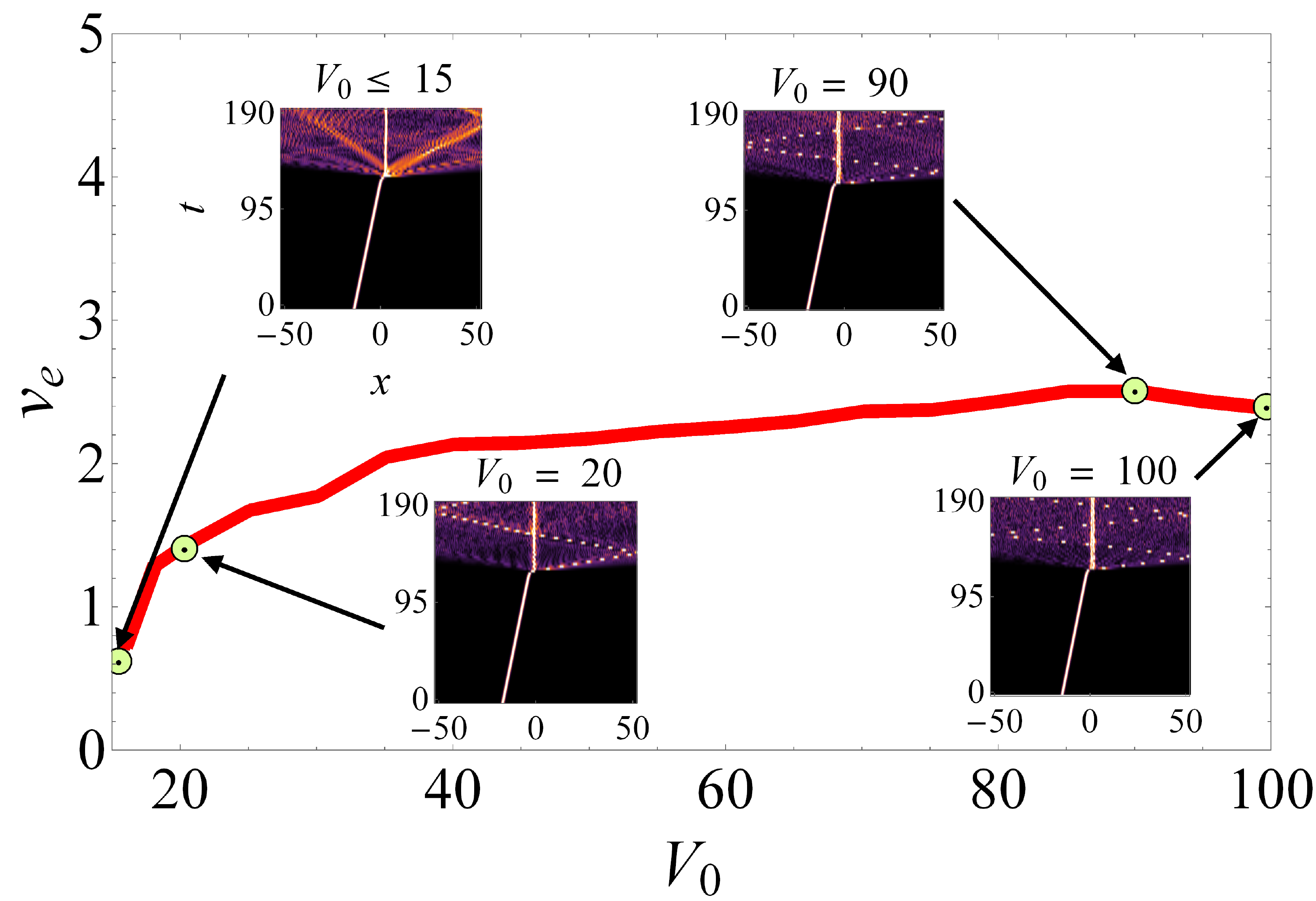}
 \caption{Speed of the ejected soliton for different potential depths. Spatio-temporal inset figures describe the dynamics of the soliton propagated with certain potential depths. Trajectory of ejected soliton appears dotted due to the high soliton speed which misses some snapshot frames. Parameters: $x_0 = -15$, $u_0 = 5$, $v_i = 0.1$, $g = 1$,  $\alpha = 3.5\sqrt{V_0}$. }
  \label{fig7}
\end{figure*}

In this section, soliton ejection mechanism of the system is characterised by varying the depth of the potential well. The depth of the potential is varied up to $V_0=100$ with $u_0 = 5$ and $v_i = 0.1$ while preserving the values of the other parameters as in the cases considered previously. The present parameter setting does not support soliton ejection for the potential depth $V_0 \leq 15$ which instead exhibits soliton trapping with a considerable amount of radiation.  Soliton ejection is initiated for $V_0>15$, as shown in Fig. \ref{fig7}. The inset plots display the soliton trapping and ejection for the corresponding values of $V_0$. For $V_0 = 16$, the speed of the ejected soliton is found to be $v_e=0.75$. For further increase in $V_0$, the ejection speed increases gradually reaching a maximum value of around $v_e=2.5$ at $V_0 = 90$. This provides a speed gain of $v_e/v_i=25$. For larger values of $V_0$, the speed of the ejected soliton reduces a little and saturates thereafter.

\subsection{Characterisation summary}
\begin{figure}[bt]\centering
\includegraphics[width=4.25cm,clip]{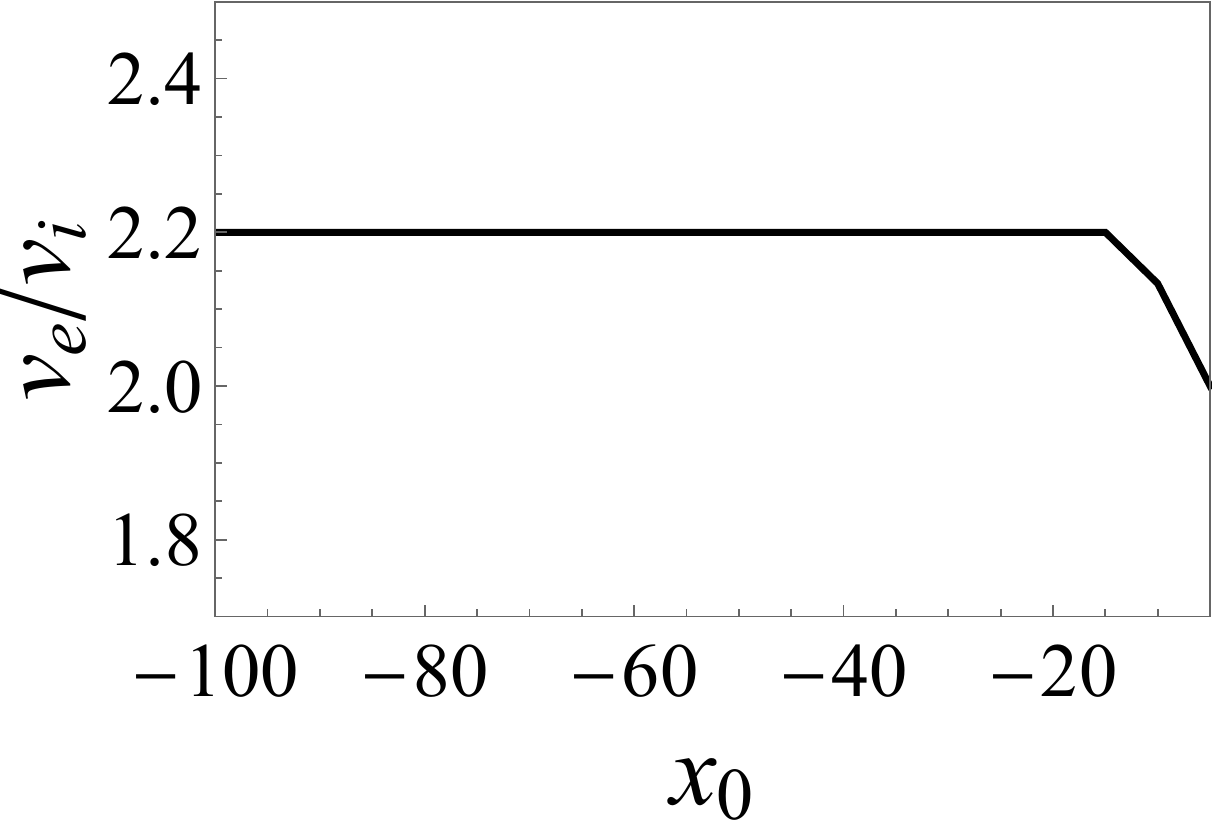}
\includegraphics[width=4.25cm,clip]{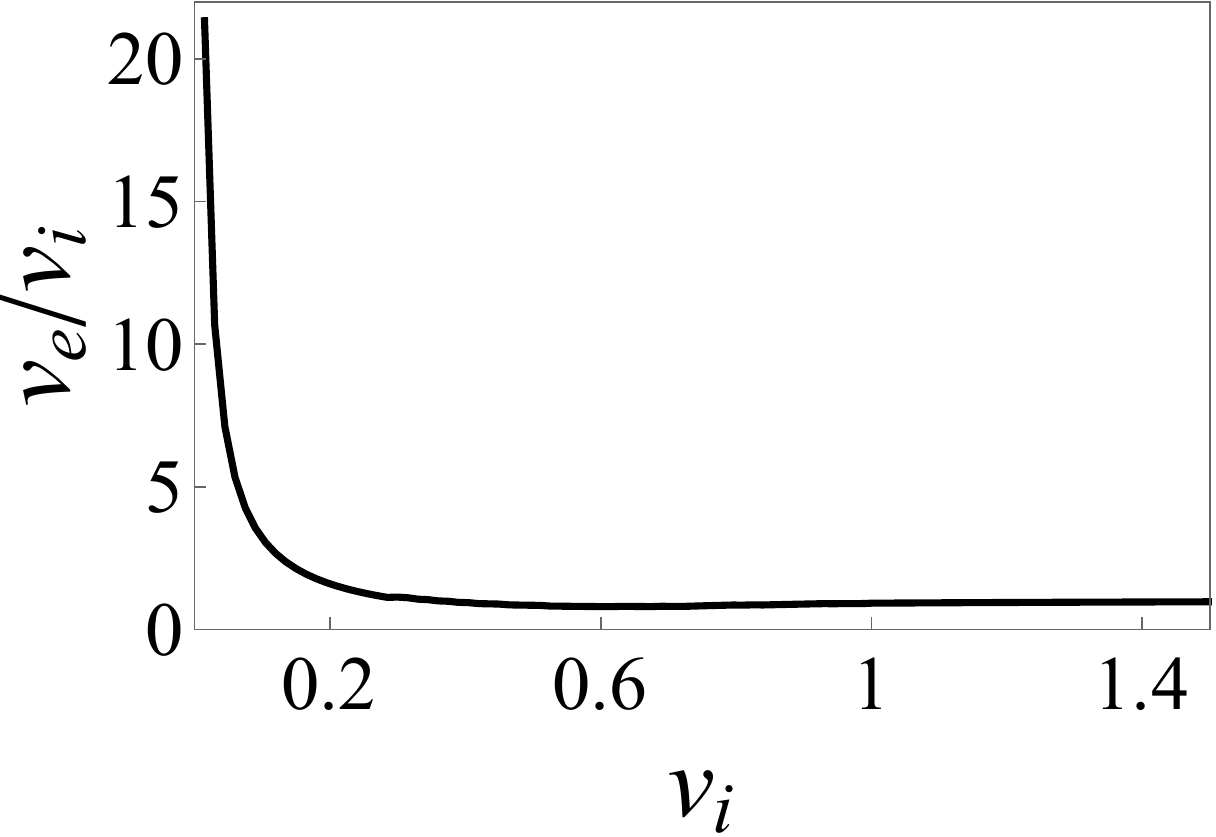}
\includegraphics[width=4.18cm,clip]{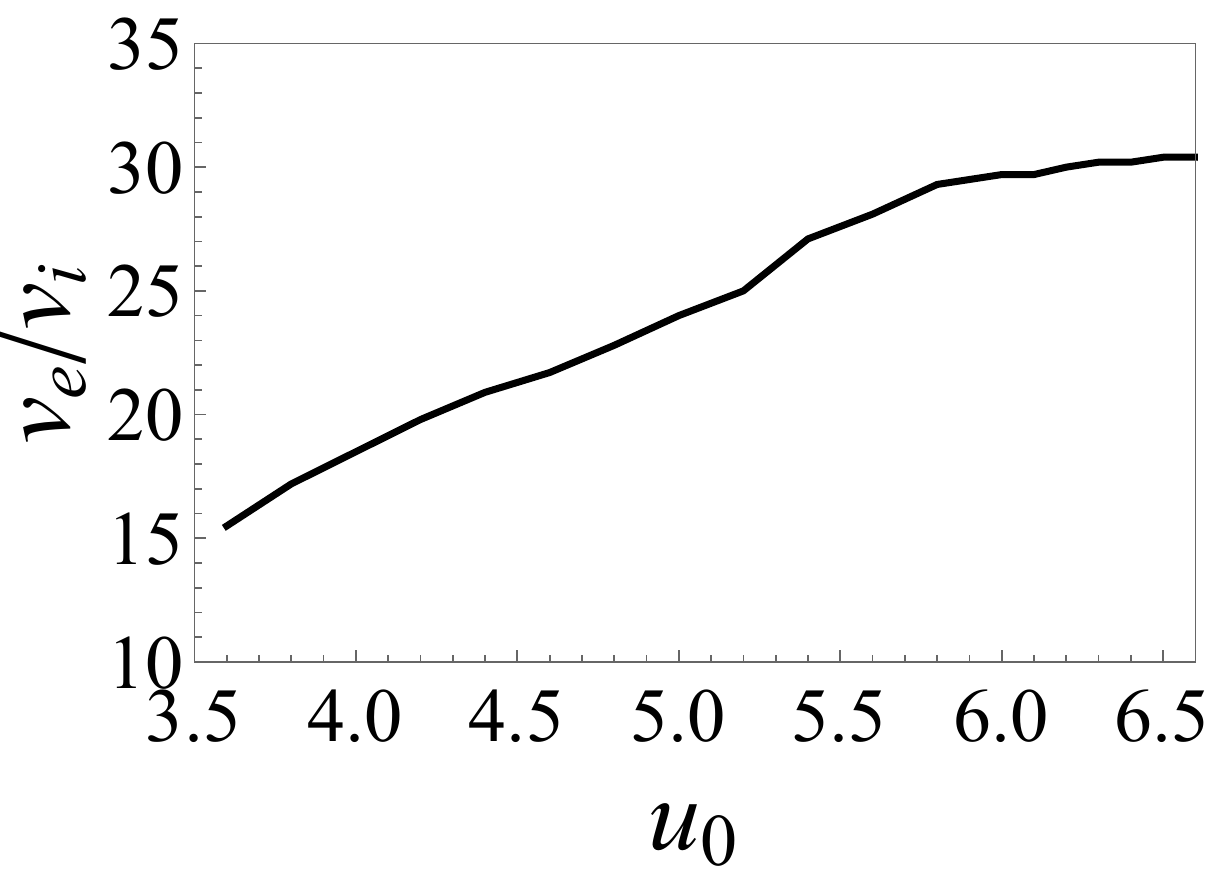}
\includegraphics[width=4.35cm,clip]{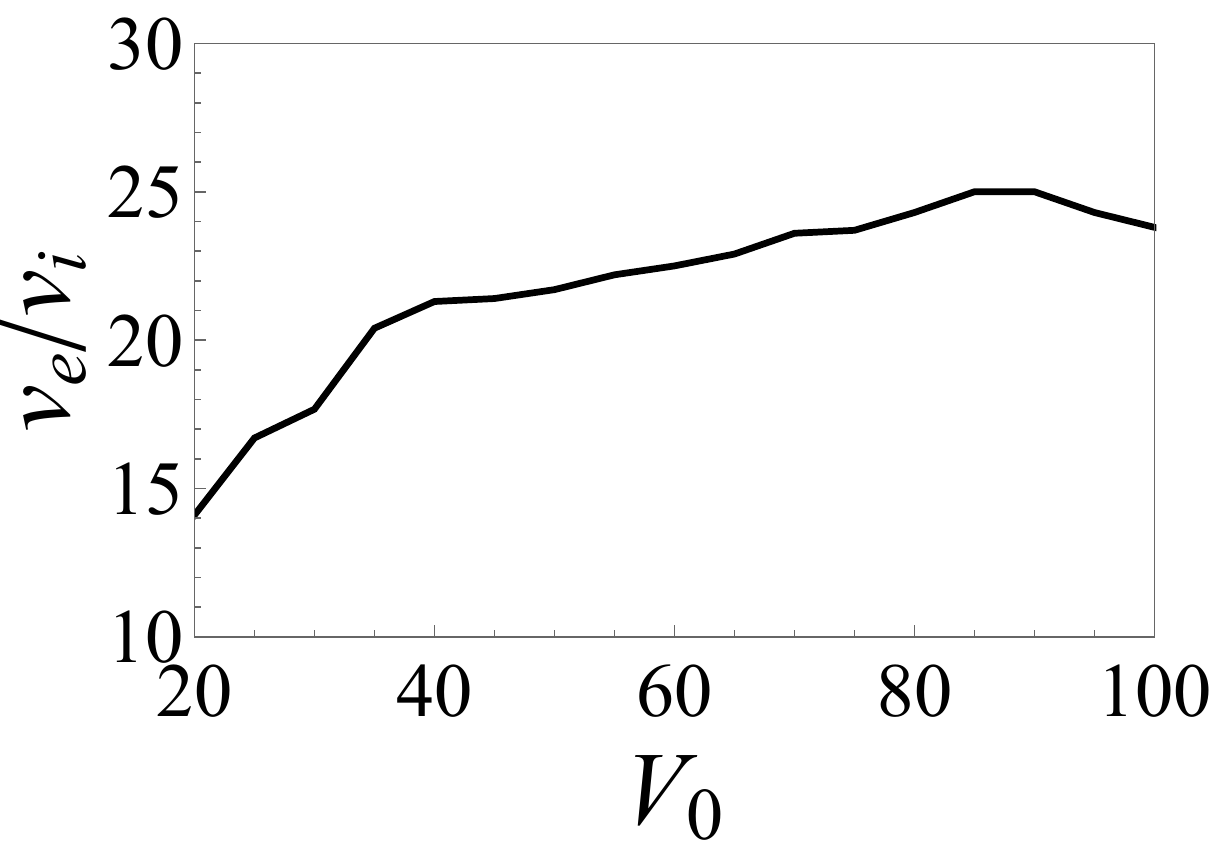}
	 \begin{picture}(5,5)(5,5)
	\put(-4.5,129.5) {\color{black}{{\fcolorbox{white}{white}{\textbf{(a)}}}}}
\end{picture}
	 \begin{picture}(5,5)(5,5)
	\put(111,172) {\color{black}{{\fcolorbox{white}{white}{\textbf{(b)}}}}}
\end{picture}
	 \begin{picture}(5,5)(5,5)
	\put(-24,43) {\color{black}{{\fcolorbox{white}{white}{\textbf{(c)}}}}}
\end{picture}
	 \begin{picture}(5,5)(5,5)
	\put(88.8,42.9) {\color{black}{{\fcolorbox{white}{white}{\textbf{(d)}}}}}
\end{picture}
 \caption{Speed gain for the different cases considered during the characterisation in terms of: (a) initial position of incident soliton, (b) initial speed of incident soliton, (c) amplitude of incident soliton, and (d) potential depth. Parameters used are the same of those in (a) Fig. (\ref{fig5}), (b) Fig. (\ref{fig2}), (c) Fig. (\ref{fig6}), (d) Fig. (\ref{fig7}).}
  \label{fig8}
\end{figure}

As a result of the above characterisation investigations, we compare the speed gain of ejected soliton for different parameter settings, as shown in Fig. \ref{fig8}, where speed gain is defined by the ratio of $v_e/v_i$. For the case of Fig. \ref{fig8}(a), the initial soliton propagated with a finite constant $v_i$ has shown a maximum gain of 2.19 times that of $v_i$ for a wide range of initial positions. However, we observe from Fig. \ref{fig8}(b), that increasing $v_i$ above a certain threshold does not yield further gain and it tends to saturate. This emphasises that lower speeds allow for the sufficiently high gain values. On the other hand, examining the speed gain in terms of input amplitude, in Fig. \ref{fig8}(c), and depth of the potential well, in Fig. \ref{fig8}(d), shed light on the possibility of achieving greater soliton ejection speeds.

In conclusion, we summarise the main results of the present section as follows. We found speed gain to be constant and not influenced by the initial position of propagation unless the input soliton approaches very closely the potential well where trapping dominates. Increasing the incident soliton speed does not yield higher gain after a certain value, where thereafter it tends to saturate. A sufficiently high ejection speed can be obtained for the incident soliton with a high amplitude and with a large potential depth, albeit the higher the incident soliton amplitude above certain threshold results in soliton ejection with considerably high radiation.

\section{Theoretical model for ejection mechanism and ejection speed}
\label{theosec}
In order to develop a theoretical model that explains the physics underlying the ejection mechanism and the ejection speed, we invoke a specific case of integrable NLSE with P\"oschl-Teller potential. For the potential
\begin{equation}
V(x)=-V_0\,{\rm sech}^2(\sqrt{2V_0}\,x),
\label{intpot}
\end{equation}
the NLSE, (\ref{nlse}), admits the exact bright soliton solution
\begin{equation}
\psi_{trap}(x,t)= \sqrt{  \frac{V_0}{g}  }\,\,{\rm sech}\left(\sqrt{2V_0}\,x\right)\,e^{iV_0t}.
\label{psitrapexact}
\end{equation}
We conjecture that this exact solution is the trapped mode left after soliton ejection. The energy of this trapped mode is given by the energy functional \cite{Hansen}
\begin{equation}
E=\int_{-\infty}^{\infty}\left[\frac{1}{2}\left|\frac{\partial\psi(x,t)}{\partial x}\right|^2-\frac{g}{2}|\psi(x,t)|^4+V(x)|\psi(x,t)|^2\right]dx
\label{trapen},
\end{equation}
which upon using (\ref{psitrapexact}), takes the form
\begin{equation}
E_{trap}=-\frac{2\sqrt{2}}{3g}V_0^{3/2}
\label{etrapanal}.
\end{equation}
The profile of the ejected soliton is modelled by that of a moving bright soliton, namely
\begin{eqnarray}\nonumber
\psi_{eject}(x,t)&=&\frac{n_e\sqrt{g}}{2}\,\,{\rm sech}\left[\frac{g\,n_e}{2}(x-x_e-v_e\,t)\right]\\&&\times e^{i\{v_e(x-x_e)+\frac{1}{8}[(g\,n_e)^2-4v_e^2]t\}}
\label{psieject},
\end{eqnarray}
where $x_e$, $n_e$, and $v_e$ are the ejected soliton position, norm, and speed, respectively.

\begin{figure}[H]\centering
\includegraphics[width=8cm,clip]{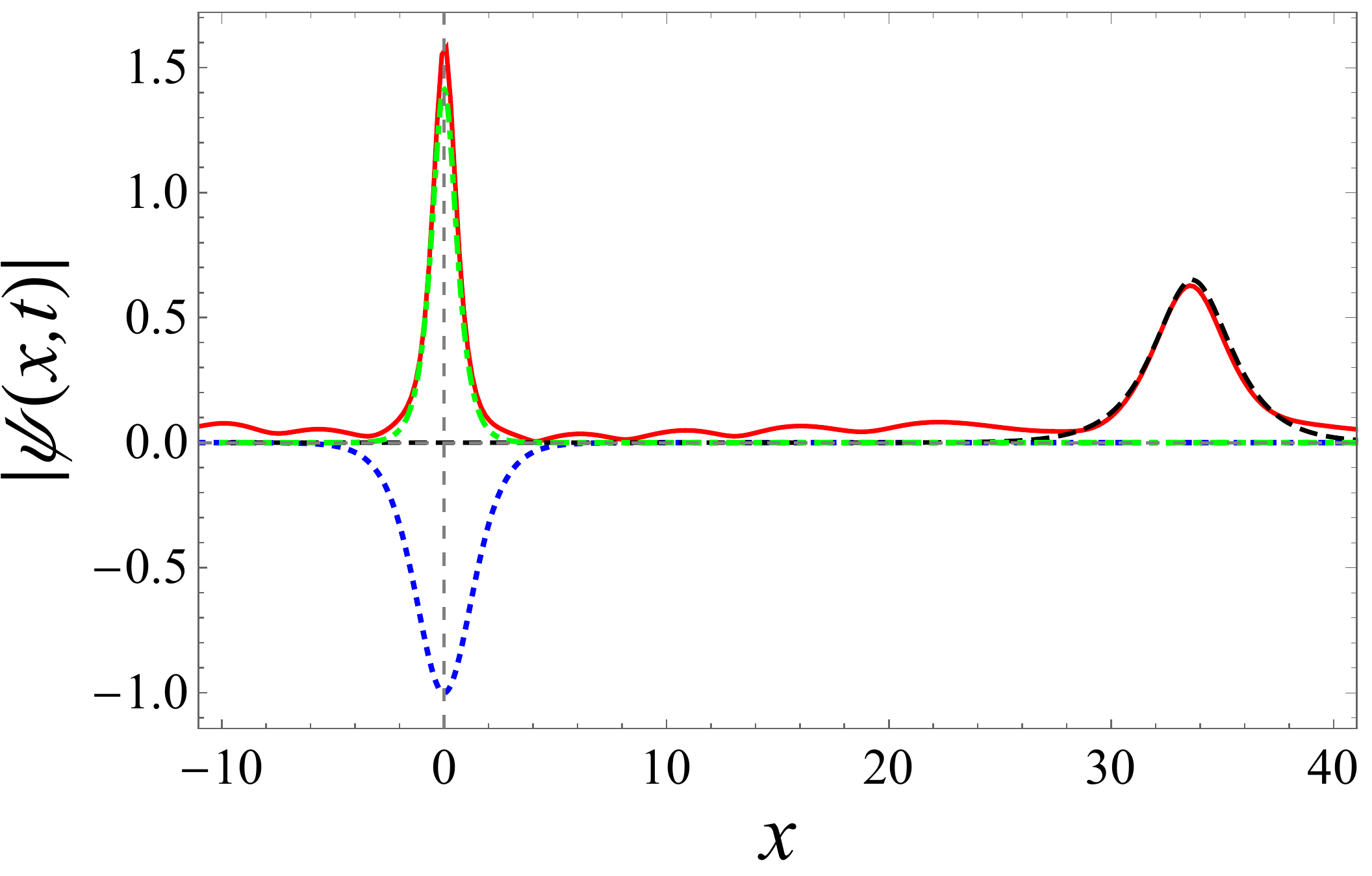}
 \caption{A snapshot of ejected and trapped soliton profiles at $t=111.49$. Solid red curve corresponds to the numerical solution of Eq. (\ref{nlse}). Dashed black curve corresponds to the analytical profile of a bright soliton, Eq. (\ref{psieject}). Dotted-dashed green curve corresponds to the exact analytical expression representing the trapped mode, Eq. (\ref{psitrapexact}). Dotted blue curve corresponds to the potential well divided by $V_0$. Parameters: $x_0 = -10$, $v_i = 0.1$, $g = 1$, $V_0 = 2$, $u_0 = \alpha = \sqrt{2V_0}$.}
  \label{fig9}
\end{figure}

\begin{figure}[H]\centering
\includegraphics[width=8cm,clip]{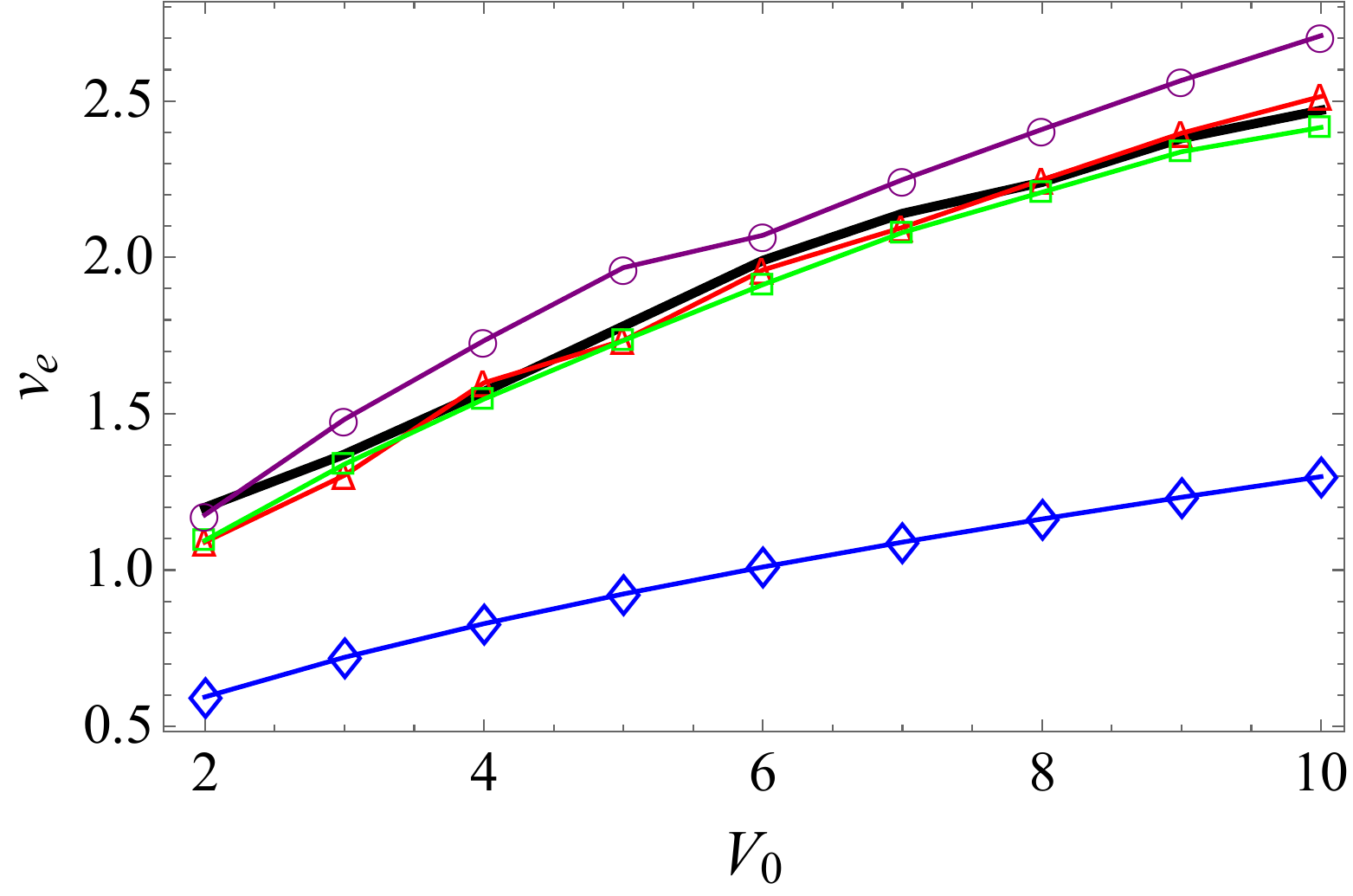}
 \caption{Ejection speed of the soliton as a function of the potential depth. Solid black curve corresponds to the numerical solution of Eq. (\ref{nlse}).  Points correspond to the ejection speed calculated using: Eq. (\ref{eject9}) (red triangles), Eq. (\ref{ve1}) (green squares),  Eq. (\ref{eject20}) (purple circles), and  Eq. (\ref{eject12}) (blue diamonds).  Parameters: $g = 1$, $v_i = 0.1$, $x_ 0 = -10$, $\alpha = u_ 0 = \sqrt {2 V_ 0}$. }\label{fig10}\end{figure}

In Fig. \ref{fig9}, a snapshot of the profiles of the ejected and trapped solitons obtained from the numerical solution of Eq. (\ref{nlse}) are plotted together with the analytical expressions Eq. (\ref{psitrapexact}) and Eq. (\ref{psieject}) for comparison. Comparison shows that the analytical bright soliton profile, Eq. (\ref{psieject}), provides an excellent representation of the ejected soliton profile. The analytical trapped soliton profile, Eq. (\ref{psitrapexact}), is also in a good agreement with the numerical profile, apart from a small deviation in the central peak. The energy of the ejected  soliton is given by inserting Eq. (\ref{psieject}) in Eq. (\ref{trapen}), and takes the form
\begin{equation}
E_{eject}=-\frac{n_e^3}{24}g^2+\frac{1}{2}n_ev_e^2
\label{ejecten},
\end{equation}
where the first term is the total of nonlinear interaction energy and the kinetic energy pressure resulting from the curvature of the soliton profile. The second term, represents the ejected soliton translational kinetic energy in terms of the ejection speed, $v_e$, which can be calculated in terms of the ejected energy as
\begin{equation}
v_e=\sqrt{\frac{2}{n_e}\left(E_{ eject}+\frac{n_e^3}{24}g^2\right)}.\label{eject9}
\end{equation}
The energy of the incident soliton is similarly given by
\begin{equation}
E_{incident}=-\frac{n_i^3}{24}g^2+\frac{1}{2}n_iv_i^2
\label{inciden}.
\end{equation}
Energy conservation relates these energies as
\begin{equation}
E_{incident}=E_{trap}+E_{eject}+E_{rad}
\label{toten1},
\end{equation}
where we have added the energy of radiation, $E_{\rm rad}$, since although it is small, neglecting it does not lead to the correct ejection speed. Substituting the incident and ejected energy expressions, Eqs. (\ref{ejecten}, \ref{inciden}) in Eq. (\ref{toten1}), the ejection speed will be given by
\begin{equation}
v_e=\sqrt{\frac{n_i}{n_e}v_i^2-2\frac{E_{trap}}{n_e}-\frac{g^2}{12n_e}\left(n_i^3-n_e^3\right)-\frac{2}{n_e}E_{rad}}
\label{ve1}.
\end{equation}
Unlike (\ref{eject9}), this expression incorporates the radiation energy, which allows for estimating its importance, as we will show next.
The condition on conservation of norm is given by
\begin{equation}
n_i=n_t+n_e+n_r
\label{normcons},
\end{equation}
where $n_r$ and $n_t$ are the norms associated with the radiation part and the trapped mode, respectively.
To estimate the effect of radiation, we neglect the radiation energy, $E_{rad}$, and radiation norm, $n_r$, in Eqs. (\ref{ve1})  and (\ref{normcons}), respectively, to get
\begin{equation}
v_e=\sqrt{\frac{12n_iv_i^2-g^2n_t(3n_i^2-3n_in_t+n_t^2)-24E_{trap}}{12(n_i-n_t)}}
\label{eject20}.
\end{equation}
Inspection, showed that a minimum amount of radiation is obtained at the resonance between the energy of the incident soliton and the energy of the trapped mode. Specifically, minimum radiation occurs when the sum of the nonlinear energy of the incident soliton and its kinetic energy pressure, namely the first term of Eq. (\ref{inciden}), is equal to the energy of the trapped mode
\begin{equation}
-\frac{n_i^3}{24}g^2=-\frac{2\sqrt{2}}{3g}V_0^{3/2},
\end{equation}
which gives
\begin{equation}
n_i=\frac{2\sqrt{2 V_0}}{g}
\label{nieq}.
\end{equation}
The norm of the trapped soliton is given by
\begin{equation}
n_t=\frac{\sqrt{2V_0}}{g}
\label{nteq},
\end{equation}
which is half the incident norm.
The ejected soliton norm can be calculated from the norm conservation condition (\ref{normcons}), and thus we are left with only the radiation norm. An approximate analytical formula for $v_e$ can thus be obtained by neglecting the radiation norm and energy
\begin{equation}
v_e=\sqrt{2v_i^2+\frac{V_0}{6}}\label{eject12}.
\end{equation}
Equations (\ref{eject9}, \ref{ve1}, \ref{eject20}, \ref{eject12}) represent different levels of approximate formulae for $v_e$. In Eq. (\ref{eject9}), we use only the numerical values for the energy and norm of the ejected soliton. As such, this formula represents the lowest level of approximation and we expect it to be the closest to the numerical results. In Eq. (\ref{ve1}), ejection speed is calculated from the conservation of the total energy. However, the advantage of this formula is that radiation energy appears explicitly allowing for an estimate on its contribution to the value of the ejection speed. This is the case of Eq. (\ref{eject20}), where both radiation norm and energy are neglected. For the same of obtaining an analytical formula without any input from the numerical quantities, we use the condition of minimum radiation to calculate the incident norm in terms of $V_0$ and then neglect radiation norm and energy, which result in Eq. (\ref{eject12}).
In Fig.  \ref{fig10}, we plot all these approximate expressions together with the exact numerical curve. The figure shows, as expected, the best agreement with the numerical curve is obtained with the expression of Eq. (\ref{eject9}). Similar agreement is obtained with  Eq. (\ref{ve1}), where radiation is taken into account. Neglecting radiation, which corresponds to Eq. (\ref{eject20}) , shifts the curve slightly above  the numerical one, indicating that radiation has somewhat a small effect on the value of the ejection speed. The approximate analytical formula, Eq. (\ref{eject12}), is clearly far from the numerical but it has a similar dependence on $V_0$ as that of the numerical result.

\begin{figure}[H]\centering
\includegraphics[width=8cm,clip]{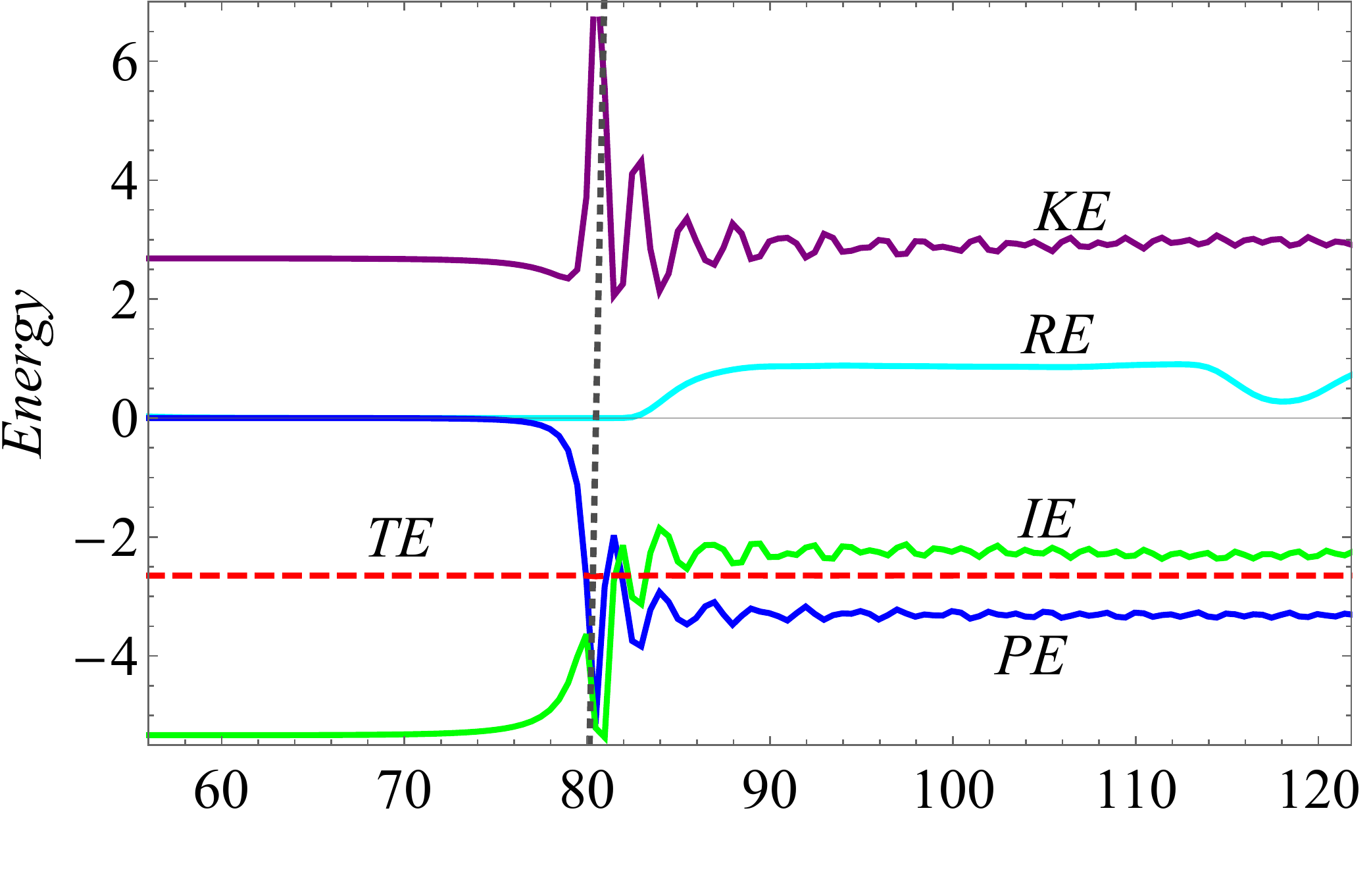}
	 \begin{picture}(5,5)(5,5)
	\put(-24,140) {\color{black}{{\fcolorbox{white}{white}{\textbf{(a)}}}}}
\end{picture}
\includegraphics[width=8.11cm,clip]{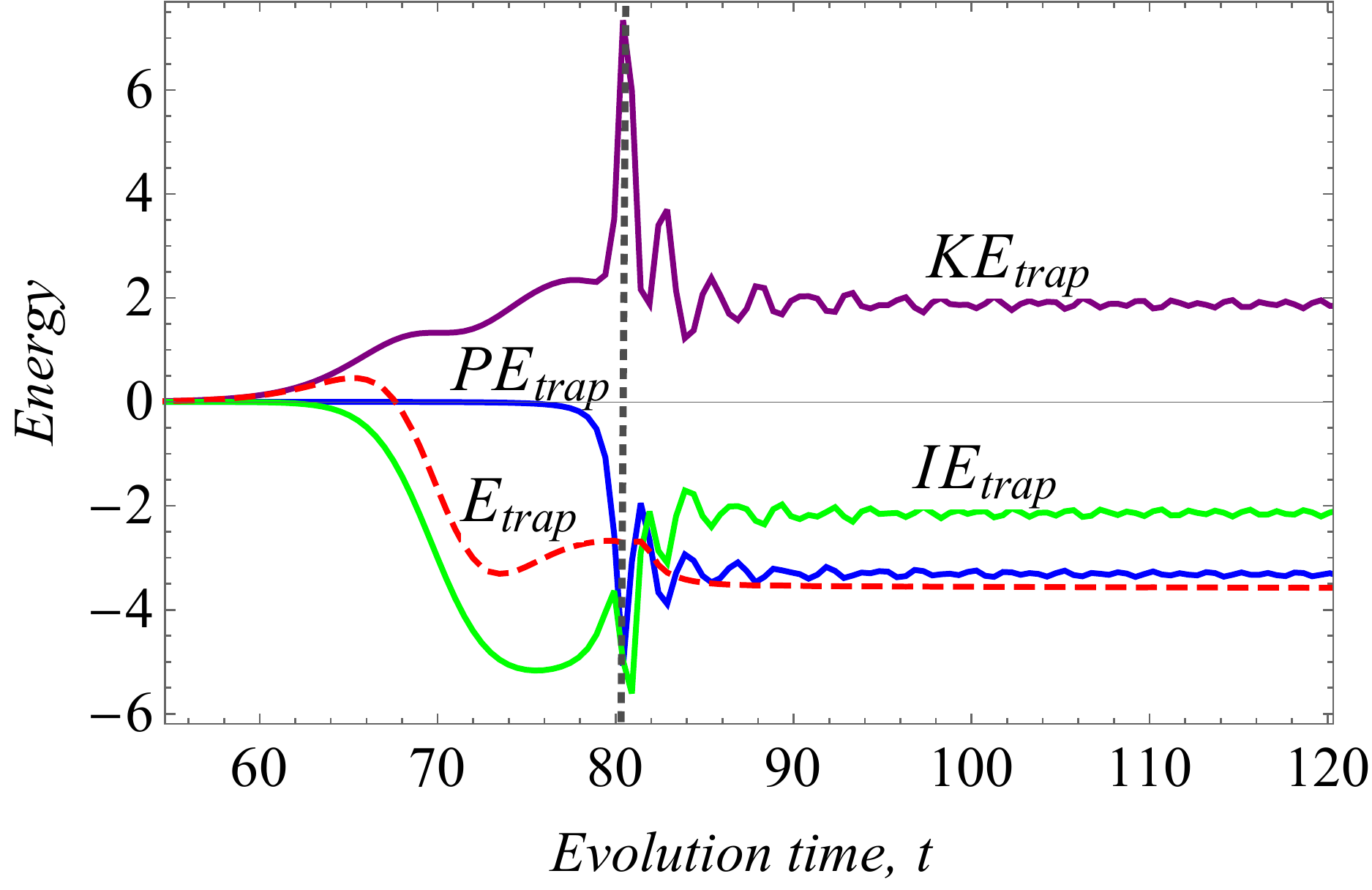}
	 \begin{picture}(5,5)(5,5)
	\put(-27,140) {\color{black}{{\fcolorbox{white}{white}{\textbf{(b)}}}}}
\end{picture}
 \caption{Time evolution of energy components for a bright soliton scattered by the potential well.  (a) total energy. (b) trapped mode energy. The curves correspond to kinetic energy ($KE$, $KE_{trap}$), interaction energy ($IE$, $IE_{trap}$), potential energy ($PE$, $PE_{trap}$), radiated energy ($RE$), and total energy ($TE$, $E_{trap}$). The point of interaction is indicated by the vertical dotted grey line. Parameters: $g = 1$, $v_i = 0.1$, $x_ 0 = -10$, $V_0 = 2$, $\alpha = u_ 0 = \sqrt {2 V_ 0}$.}
  \label{fig11}
\end{figure}
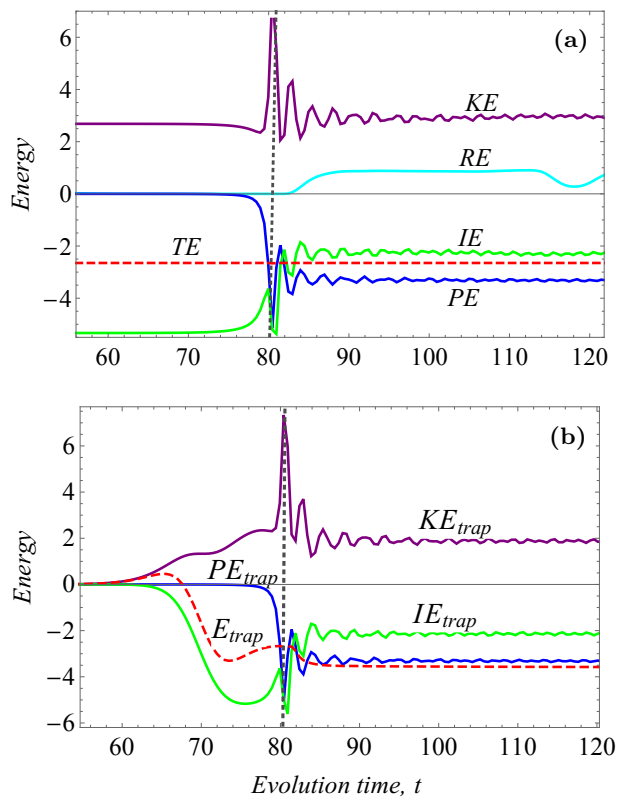

To qualitatively understand the physics underlying the ejection effect, we calculate the different energy components of the soliton during the whole evolution time interval. Figure \ref{fig11}(a) shows the total kinetic energy, total interaction  energy, total potential energy, total radiated energy, and total energy. The same quantities for the trapped mode related by $ E_{trap}={KE_{trap}}+{PE_{trap}}+{IE_{trap}}$ are plotted in Fig. \ref{fig11}(b). Clearly,  after scattering the total kinetic energy is slightly larger than that of the trapped mode. Since the trapped mode is stationary, then its kinetic energy is only in the form of kinetic energy pressure. Thus, the difference between the total kinetic energy and the trapped mode kinetic energy after scattering is  translational kinetic energy which is distributed between the ejected soliton and the radiation part. The total trapped mode energy is less than that of the incident soliton. The increase in kinetic energy compensates exactly to this difference such that the total energy is conserved.

The presence of impulsive potential energy adds to the effect of the focusing nonlinear interaction. This leads to that the norm of original soliton is too large to sustain, since the impulsive forces become larger than that of the expulsive kinetic energy pressure. Equilibrium is retained by ejecting some of the incident soliton norm. The nonlinear interaction in the ejected intensity is responsible for forming the ejected soliton profile. The difference between the energy of the trapped mode and the incident soliton is transferred to the ejected soliton and radiation. Since that difference is negative, the ejected soliton energy will be positive, i.e., translational kinetic energy.

Further insight is obtained by observing the dynamics of soliton profile immediately after it enters the potential region. We observe that the whole soliton profile is first confined by the potential with a compressed width and increased intensity. This results in increasing the kinetic energy pressure, as is clearly shown by the spike in the kinetic energy curve in Fig. \ref{fig11}. Associated with this positive spike, there is a negative spike in the potential energy occurring at the same time such that the total energy is conserved. The interaction energy is changed only slightly at this moment. While the energy is conserved for the highly compressed state, it is not a stable equilibrium and the kinetic energy pressure starts expanding the soliton. This expansion accelerates part of the soliton, which should be its outer shell, giving it translational kinetic energy to leave the original soliton as an ejection. The direction of the ejection is towards the part that had the first encounter with the potential since this part is exposed to the compression for a longer period than the opposite part.

\section{Multi-node trapped modes and two-soliton ejection}
	\label{multisec}
	In this section, we consider first soliton ejection associated with the formation of multi-node trapped modes. We investigate the relation between the ejection speed and the trapped energy of the mode. We then investigate the ejection stimulated  by two solitons  launched simultaneously from both sides of the potential.
	
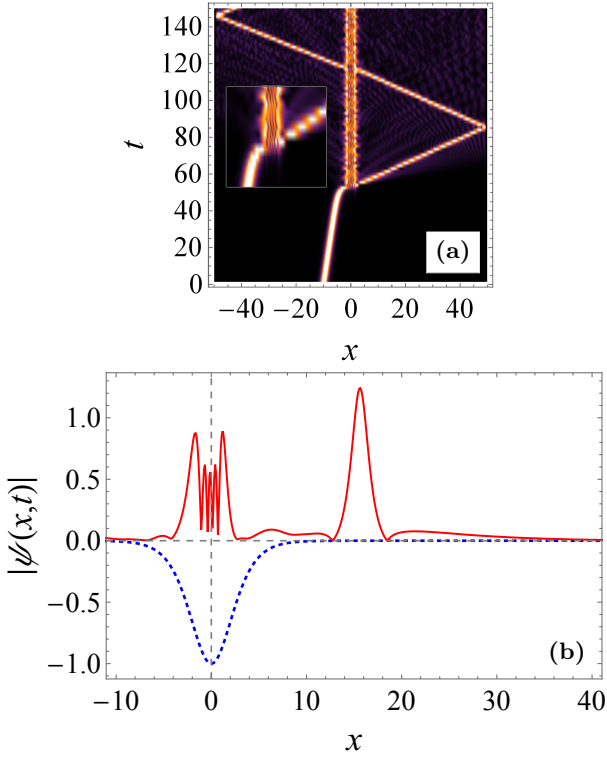
\begin{figure}[H]
	\centering
\includegraphics[width=5cm,clip]{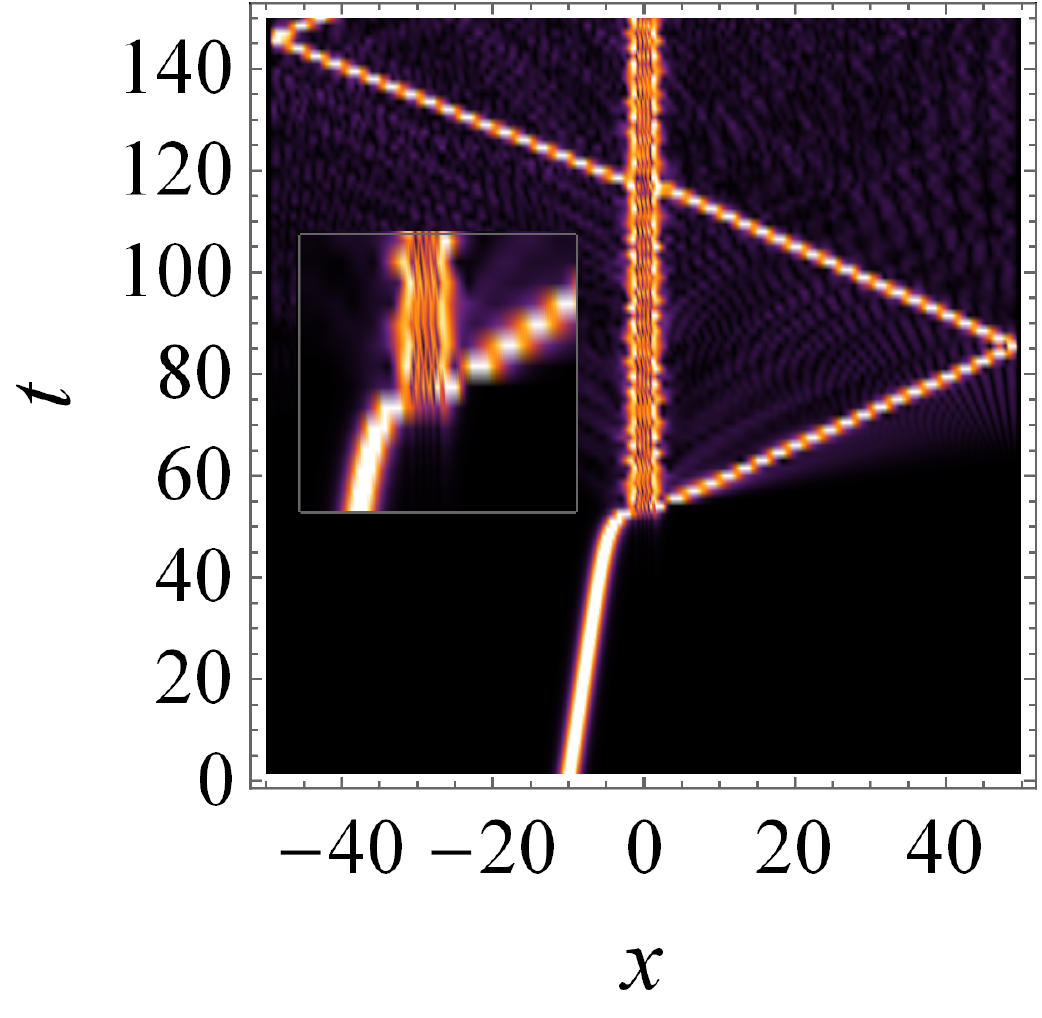}
	 \begin{picture}(5,5)(5,5)
	\put(-24,47){\color{black}{{\fcolorbox{white}{white}{\textbf{(a)}}}}}
\end{picture}
\includegraphics[width=8cm,clip]{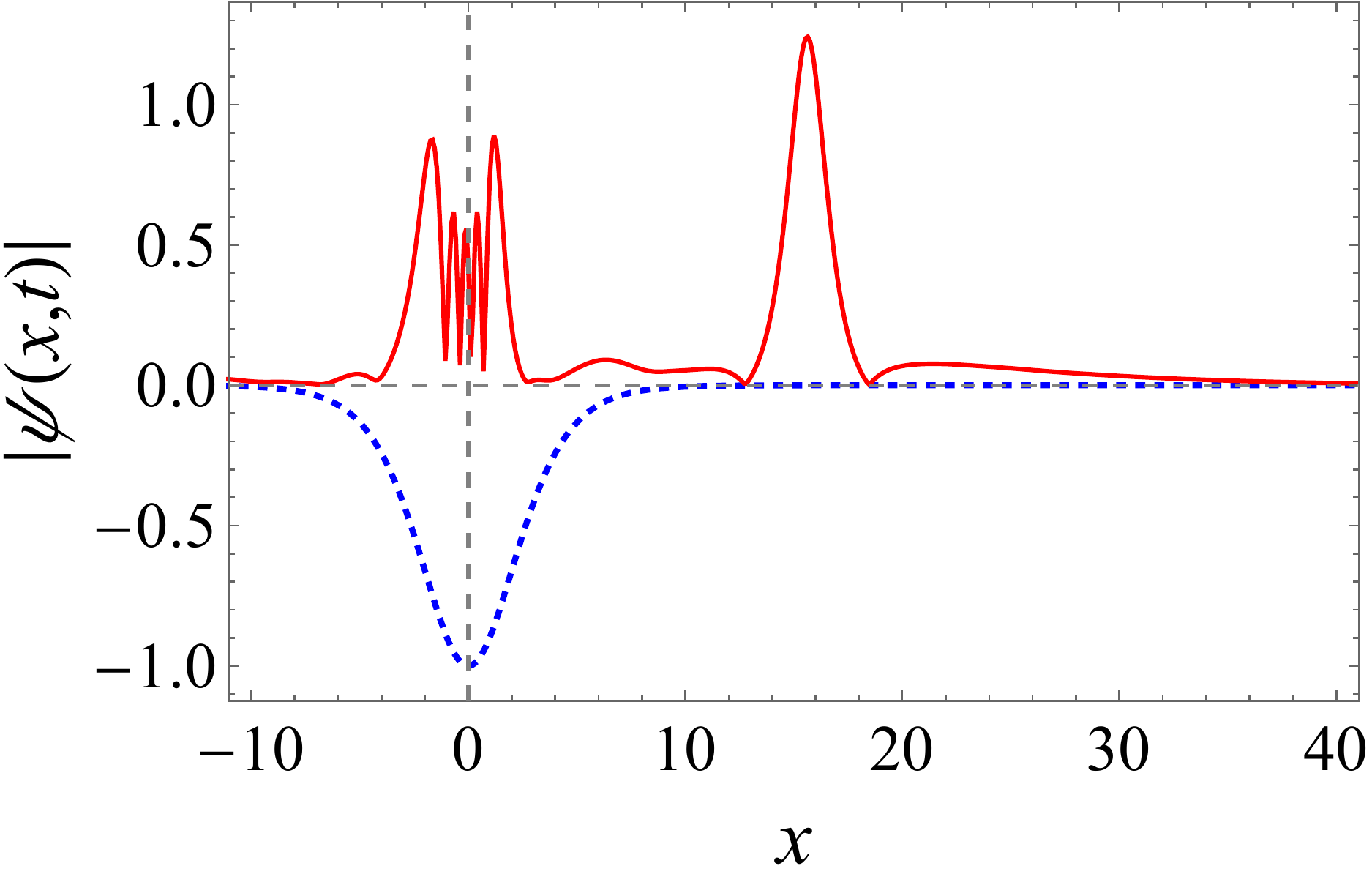}
		 \begin{picture}(5,5)(5,5)
	\put(-24,43){\color{black}{{\fcolorbox{white}{white}{\textbf{(b)}}}}}
\end{picture}
 \caption{(a) Spatio-temporal plot showing  4-node trapped mode and (b) pulse profile of the ejected soliton at $t=63$. Solid red curve corresponds to the numerical solution of Eq. (\ref{nlse}) and dotted blue curve corresponds to the potential well divided by $V_0$. Inset in (a) shows a zoom-in of the scattering near the potential region.
 Parameters: $u_0=2$, $V_0=22$,  $\alpha=\sqrt{V_0}/5$, $x_0=-10$,  $v_i=0.1$, $g=1$.}
  \label{fig12}
\end{figure}

	Based on the qualitative understanding of the mechanism of soliton ejection which we reached in the previous section, we conclude that when the magnitude of trapped mode energy is large, the ejection speed should be large. Multi-node trapped modes do indeed have binding energy magnitudes larger than the nodeless mode considered in the previous section.  To realise a situation where soliton ejection results from the trapping of a multi-node trapped mode, we increase the width of the potential in order to accommodate the multi-nodes trapped modes. In Fig. \ref{fig12}, we show such a case of 4-nodes trapped mode. The ejection speed in this case equals $v_e=1.49$. In Table \ref{table1},  we list the trapped energy and its three components, the kinetic, ${KE_{trap}}$, the potential, ${PE_{trap}}$, and the interaction, ${IE_{trap}}$, energies related by $ {E}_{trap}={KE_{trap}}+{PE_{trap}}+{IE_{trap}}$. The list is recalculated for a range of nodes and shows the corresponding ejection speed in each case. As shown in the table, larger magnitude of binding energy $E_{trap}$ is always associated with larger trapped kinetic energy  ${KE_{trap}}$. In the previous section, we have concluded that the difference between the total incident energy and the trapped kinetic energy, ${KE_{trap}}$, turns to the translational kinetic energy of the ejected soliton plus the small radiation part. This is verified here, by confirming a linear relationship between $E_{trap}$ and $v_e^2$. 
	
	Another interesting situation is the two-soliton ejection stimulated by simultaneous scattering of two solitons with the potential from both of its sides. The incident two-soliton profile considered here, is given by
\begin{eqnarray}\label{bright}
	\psi(x,t)&=&\psi_+(x,t)+\psi_-(x,t)e^{i\Delta \phi},
\end{eqnarray}
\begin{eqnarray}\nonumber
	\psi_{\pm}(x,t)&=&\frac{u_0}{\sqrt{g}}\,\text{sech}\left[u_0\left(\pm v_it-x\pm x_0\right)\right]\\&&\times e^{\frac{i}{2}\left\{u_0^2t-\left[v_i\left(v_it\mp 2x+2x_0\right)\right]\right\}},
\end{eqnarray}
	where $\psi_+(x,t)$ is the soliton launched from the left and is moving to the right, and $\psi_-(x,t)$ is the soliton launched from the right and moving to the left. For both cases, we assume $x_0<0$ and  $v_i>0$. The relative phase difference between the two incoming solitons is  $\Delta \phi\in[0,2\pi]$.  
	
\begin{figure}[H]\centering
\includegraphics[width=4cm,clip]{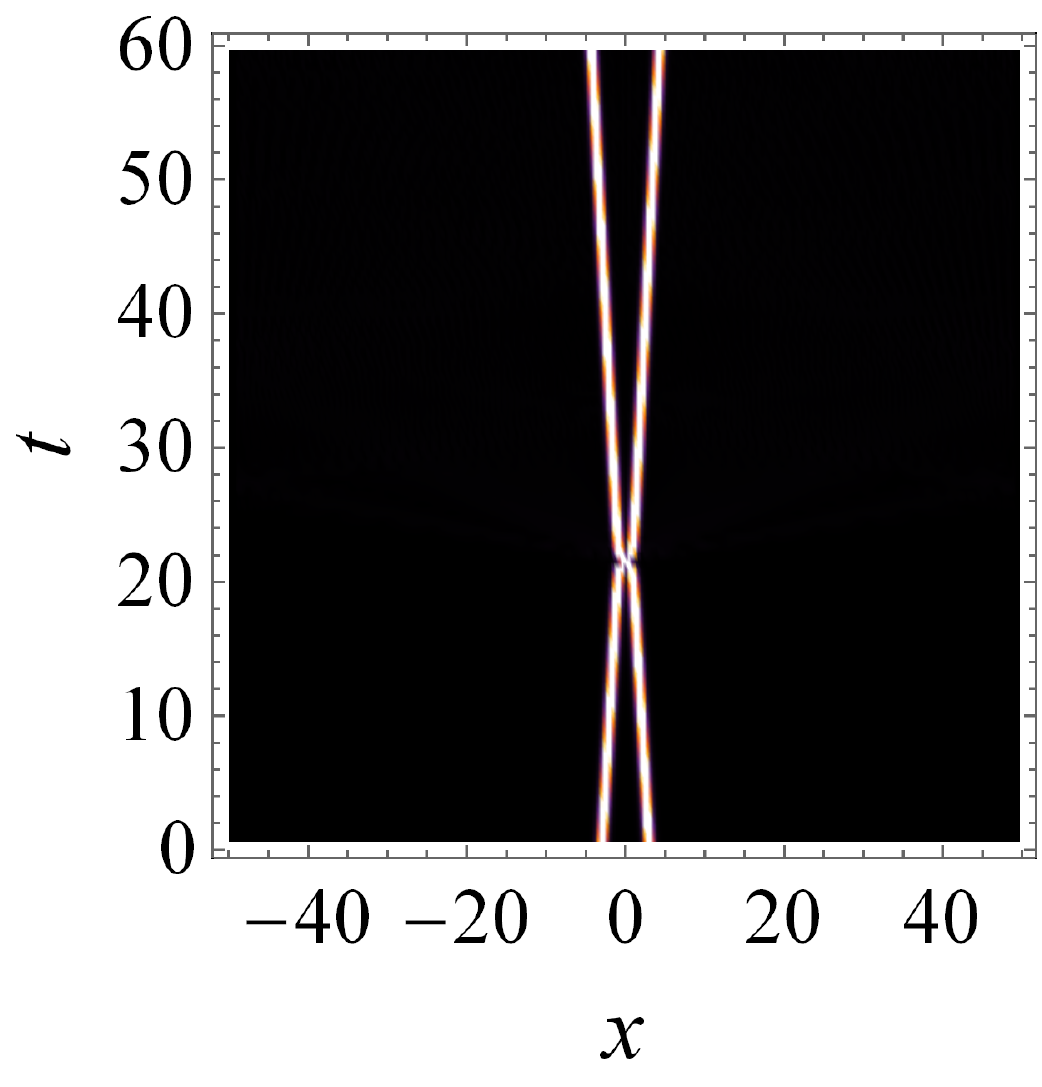}
	 \begin{picture}(5,5)(5,5)
	\put(-23,40) {\color{black}{{\fcolorbox{white}{white}{\textbf{(a)}}}}}
\end{picture}
\includegraphics[width=4cm,clip]{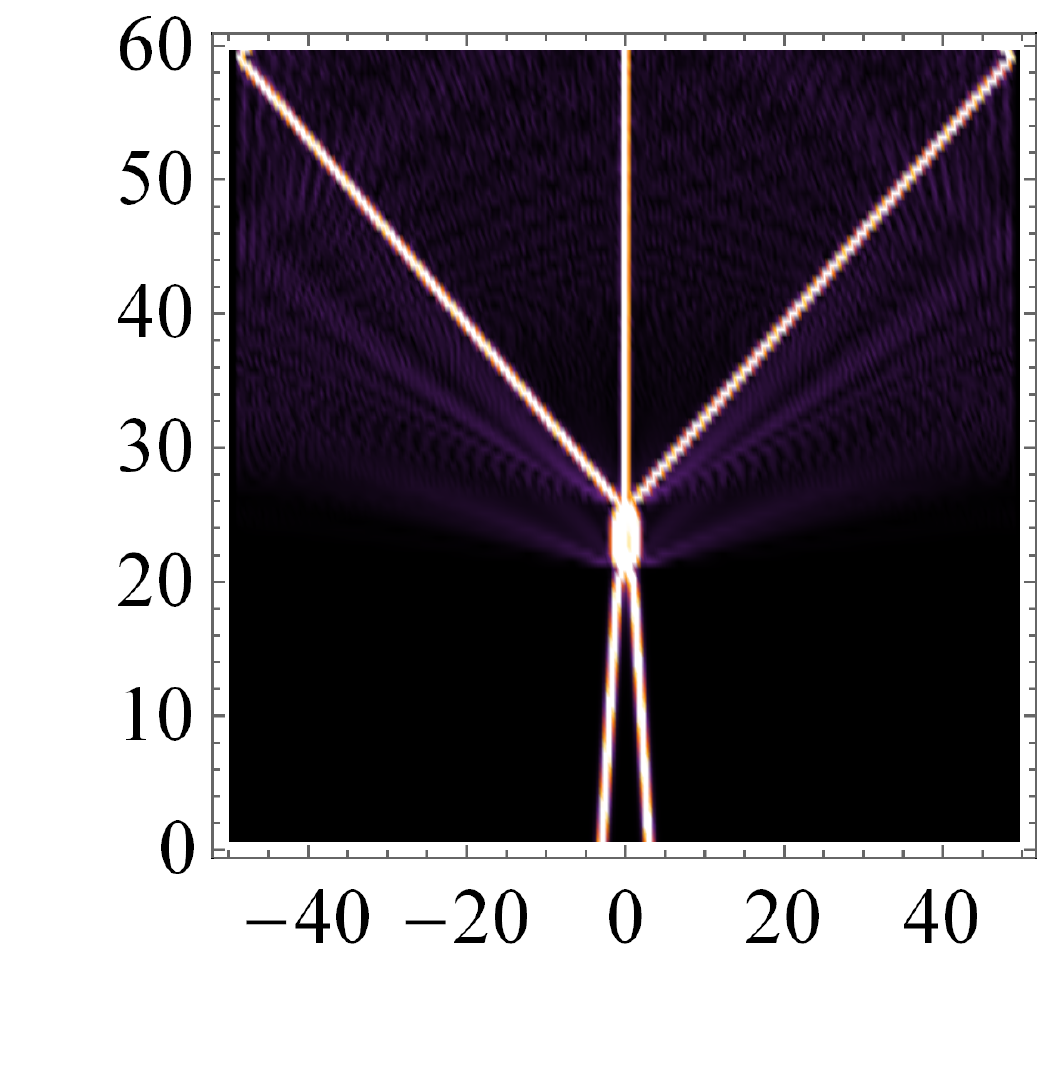}
	 \begin{picture}(5,5)(5,5)
	\put(-23.5,40) {\color{black}{{\fcolorbox{white}{white}{\textbf{(b)}}}}}
\end{picture}
\includegraphics[width=4cm,clip]{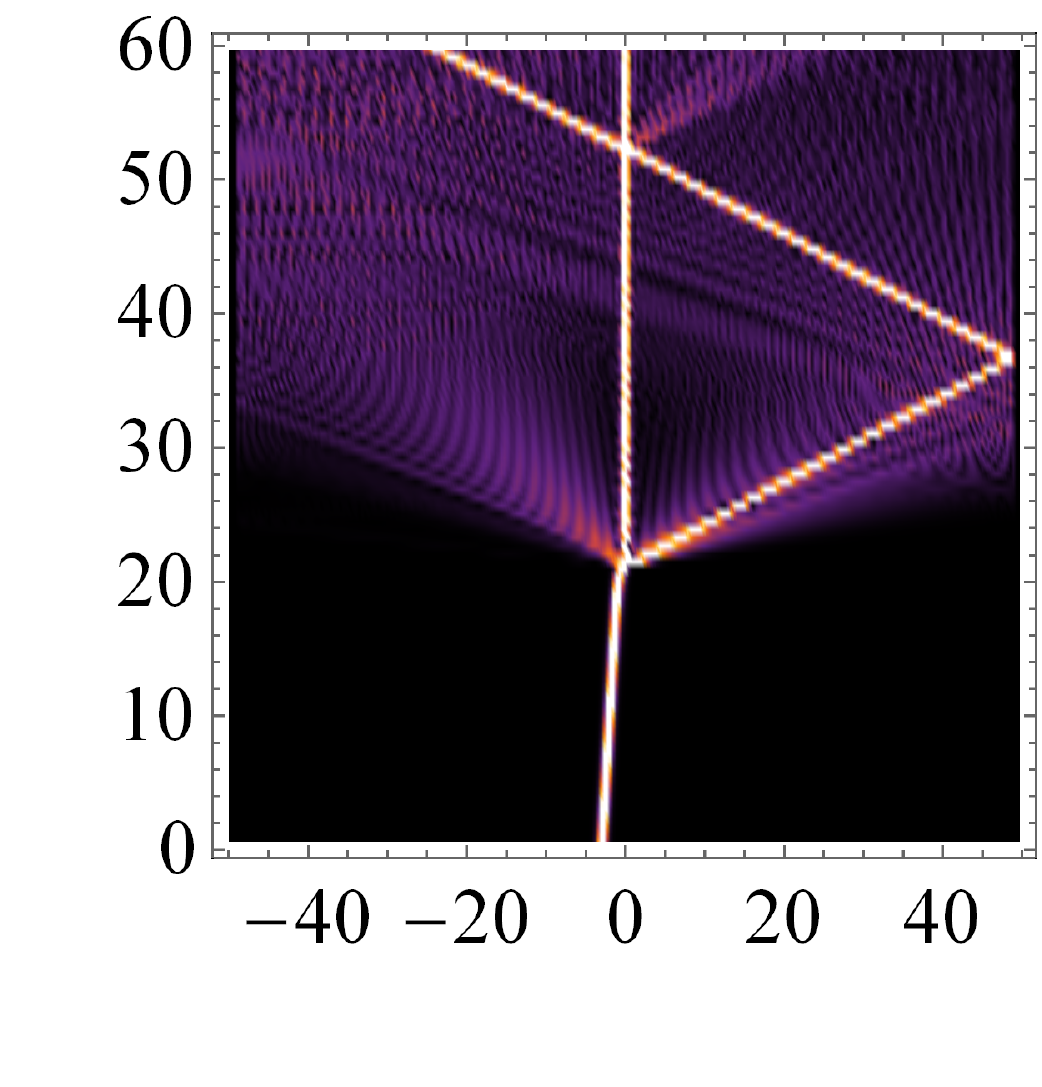}
	 \begin{picture}(5,5)(5,5)
	\put(-23,40) {\color{black}{{\fcolorbox{white}{white}{\textbf{(c)}}}}}
\end{picture}
\includegraphics[width=4cm,clip]{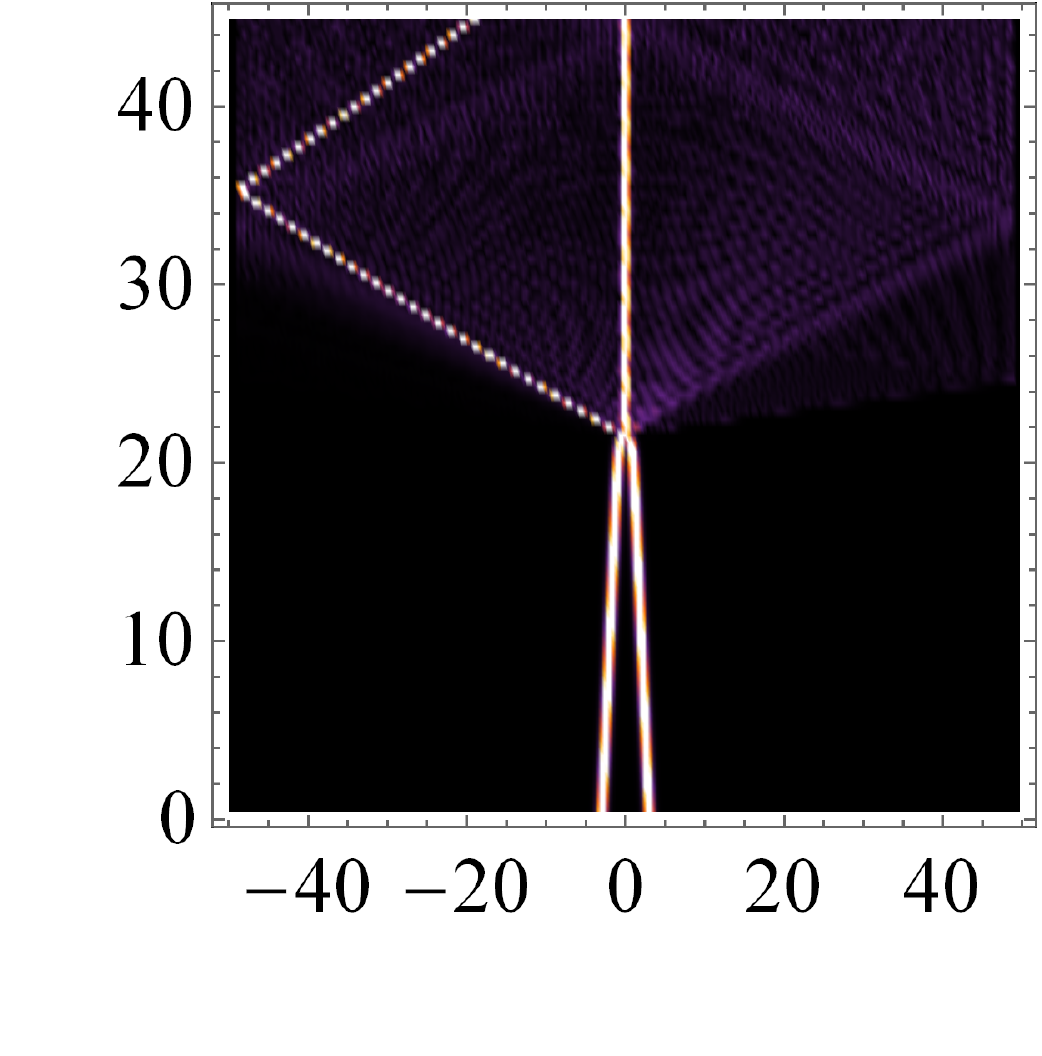}
	 \begin{picture}(5,5)(5,5)
	\put(-23.5,40) {\color{black}{{\fcolorbox{white}{white}{\textbf{(d)}}}}}
\end{picture}
 \caption{Two solitons scattered by the potential from both of its sides. (a) two out-of-phase solitons, $\Delta\phi=\pi$. (b) two in-phase solitons, $\Delta\phi=0$, with a symmetric soliton ejection. Ejection speed is $v_e=0.43$. (c) single-soliton ejection with ejection speed $v_e=3.40$. (d) asymmetric soliton ejection for $\Delta\phi=0.9\,\pi$.  Ejection speed is $v_e=3.9$.  Parameters: $u_0=\sqrt{2\,V_0}$, $V_0=16$, $\alpha=\sqrt{2\,V_0}$, $x_0=-3$, $v_i=0.1$,  $g=1$.}
  \label{fig13}
\end{figure}
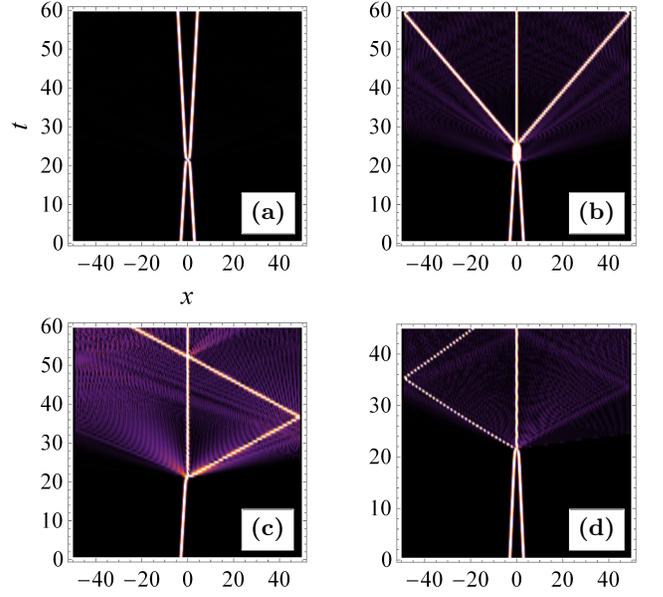

	Here, we are specifically interested in investigating the effect of the relative phase difference between the two incoming solitons on the ejected solitons.  We find that symmetric two-soliton ejection occurs always when $\Delta \phi$ is between the two special cases of in-phase and out-of-phase values. Distinguished behaviours take place when the solitons are exactly in-phase or out-of-phase. In the following, we consider in particular three cases of the relative phase difference, $\Delta \phi=0,\,0.9\,\pi,\,\pi$. The results turn out to be as follows: i)  when the two solitons are exactly out-of-phase, namely $\Delta \phi=\pi$,  no soliton ejection takes place. The two solitons interfere destructively at the potential region which results in an effective repulsive force, as it is known for two-soliton scattering \cite{two}. Consequently, no trapped mode forms and the two solitons are fully reflected with the same but opposite speeds since all energy carried by the two incoming solitons will be transferred completely to the two reflected solitons. This case is shown  in Fig. \ref{fig13}(a), ii)  when the two solitons are exactly in-phase, namely $\Delta \phi=0$, they will interfere constructively and a trapped mode forms. Figure \ref{fig13}(b) shows this situation. Since the resulting trapped mode energy and its norm are mainly set by the parameters of the potential, the energy difference  between the trapped mode and the incident wave will be transferred to the ejected solitons. A symmetric soliton ejection forms such that each ejected soliton will have the same norm,  ${n_{eject}}_{_l}={n_{eject}}_{_r}=7.74$ and carry equal ejected energies, ${E_{eject}}_{_l}={E_{eject}}_{_r}=-18.5$. Ejection speed in this case, $v_e=1.51$, is almost half that of a single-soliton ejection, $v_e=3.40$, attained by the same parameters shown in Fig. \ref{fig13}(c). iii)   when the relative phase difference is any value other than the out-of-phase or in-phase cases, for example $\Delta \phi=0.9\,\pi$, an asymmetric soliton ejection occurs, as shown in Fig. \ref{fig13}(d). Due to a partially destructive interference, only one soliton will eject with a considerably higher speed, namely $v_e=3.9$. The energy distributed over two ejected solitons for the in-phase case, is now transferred to one ejected soliton which has almost twice the norm as one of the two ejected solitons, and its kinetic energy will be almost doubled leading to an increased ejection speed. This can be induced from the trapped energy in the two cases shown  in Table \ref{table2}, were we have summarised the norms and energies of the two cases associated with $\Delta \phi=0,\,0.9\,\pi$.
	 To verify this understanding, we list in Table \ref{table3}  the  energy components of the trapped energy of the two cases. The trapped kinetic energy in the case of $\Delta \phi=0.9\,\pi$ equals ${KE_{trap}}=57.4$ which is higher than that for the case of two in-phase solitons, ${KE_{trap}}=44.1$. This conclusion is consistent with the situation of single-soliton ejection with multi-node modes discussed at the begging of this section.
	
	 To affirm the solitonic nature of the ejected pulse, we calculate the ejection speed using the energy expression of the ejected soliton, which we denote by ${{v_{e}}_{_{theo}}}$, namely $E_{eject}=-\frac{n_e^3}{24}+\frac{1}{2}n_e{{v_{e}}_{_{theo}}}^2$, for $g=1$, with the numerical values of $E_{eject}$ and $n_e$.  As shown in the last column of Table \ref{table2}, the calculated value agrees well with the numerical value ${{v_{e}}_{_{num}}}$.
	
\begin{figure}[H]\centering
\includegraphics[width=5cm,clip]{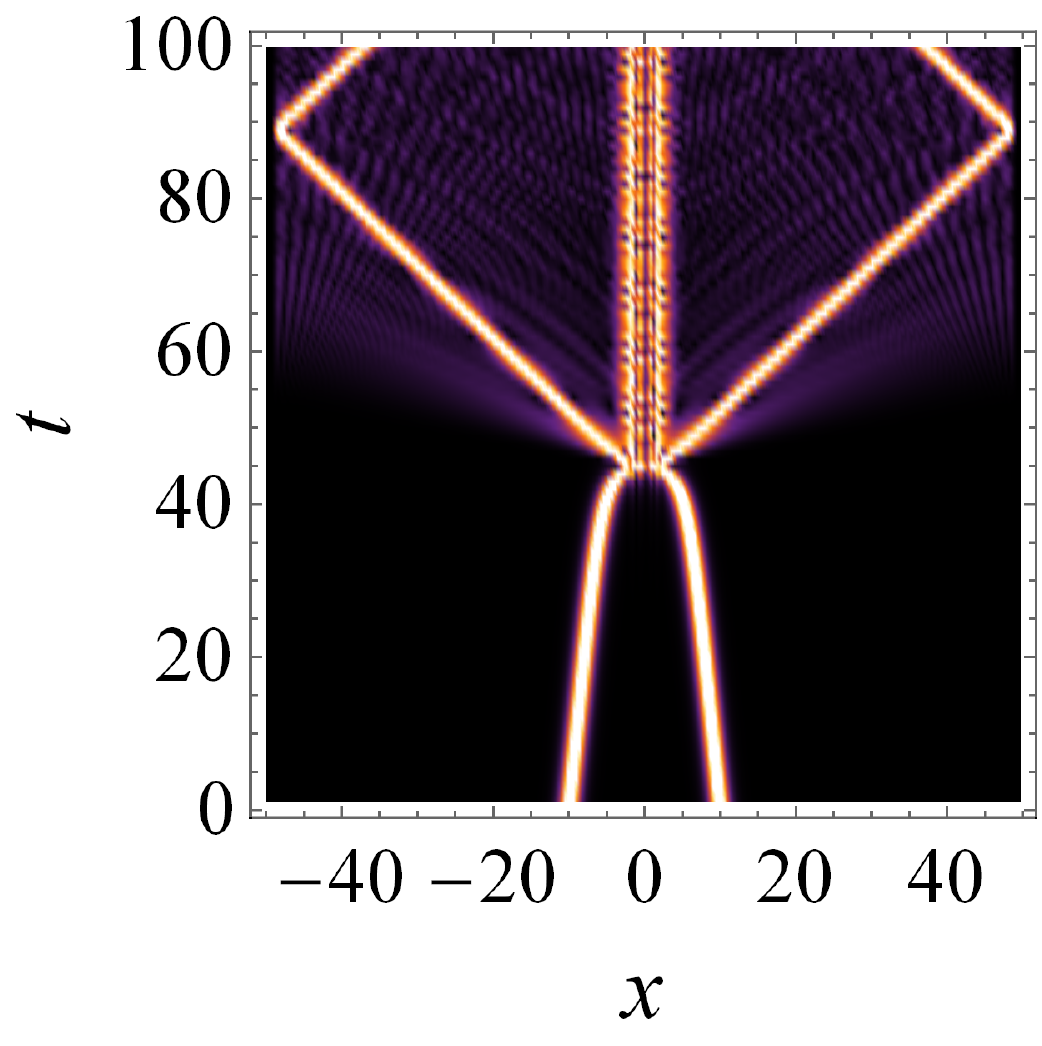}
 \caption{Symmetric two soliton ejection resulting from two out-of-phase input solitons,  $\Delta \phi=\pi$. Ejection speed is $v_e=1.12$.   Parameters: $u_0=2$, $V_0=22$, $\alpha=\sqrt{V_0}/6$, $x_0=-10$, $v_i=0.1$,  $g=1$.}
  \label{fig14}
\end{figure}

	In addition to the above findings, changing the depth and width of the potential allows for  other interesting two-soliton ejection phenomena. As an example, we show in Fig. \ref{fig14} two out-of-phase solitons form a multi-node trapped mode and eject symmetrically with ejection speed is $v_e=1.12$.
	
	 In view of the above, it is possible to design a nonlinear soliton interferometry which can be used as a detector of the  phase difference between solitons.
	
\begin{table}[h]
	\centering
	\begin{math}
		\begin{array}{|ccccccc|}
			\hline
		\sqrt{V_0}/\alpha& {\rm no.\,of\, nodes} & E_{trap}&{KE_{trap}}& {PE_{trap}}&{IE_{trap}}& v_e  \\
			\hline
			3 & 2 & -2.39&4.47&-6.53&-0.33&0.538  \\
				4 & 3 &-4.25&6.26&-10.21&-0.292&1.220 \\
			5 & 4 &-5.10&8.55&-13.17&-0.488&1.490  \\
			6 & 5 &  -5.82&10.17&-15.61&-0.37&1.690\\
			7 & 6 & -6.38&12.14&-18.24&-0.24&1.817\\
			8 & 7 & -6.91&13.61&-20.01&-0.51&1.940\\
		9 & 8 &-7.33&16.48&-23.51&-0.31&2.060 \\
			10 & 9 &-8.00&21.66&-29.30&-0.36&2.160
		\\
			\hline
		\end{array}
	\end{math}
	\caption{Ejection speed for multi-node trapped modes.  Parameters: $u_0=2$, $V_0=22$, $x_0=-10$, $v_i=0.1$, $g=1$.}
	\label{table1}
\end{table}

\begin{table*}[bt]
	\centering
	\begin{math}
		\begin{array}{|ccccccccccccc|}
			\hline
			\Delta \phi& n_{tot} &n_{trap}& {n_{eject}}_{_l}&{n_{eject}}_{_r}& n_{rad} &TE& E_{trap}&{E_{eject}}_{_l}&{n_{eject}}_{_r}&E_{rad}&{{v_{e}}_{_{num}}}&{{v_{e}}_{_{theo}}}  \\
			\hline
			0&22.8&7.1&7.74&7.74&0.1&-120.835&-84&-18.5&-18.5&0.97&0.43&0.46\\
			0.9\pi&22.8&8.4&14.1&0&0.25&-120.835&-110&-12.3&0.35&0.56&3.9&3.85 \\
			\hline
		\end{array}
	\end{math}
	\caption{The norms, energies, and ejection speeds associated with the dynamics of two-soliton ejection by two in-phase solitons, $\Delta \phi=0$, and two solitons with a phase difference  $\Delta \phi=0.9\pi$. The last two columns correspond to the numerical value of ejection speed, ${{v_{e}}_{_{num}}}$, and the calculated one from the theoretical model, ${{v_{e}}_{_{theo}}}$. Parameters: $u_0=\sqrt{2\,V_0}$, $V_0=16$, $\alpha=\sqrt{2\,V_0}$, $x_0=-3$, $v_i=0.1$,  $g=1$.}
	\label{table2}
\end{table*}
\begin{table}[bt]
	\centering
	\begin{math}
		\begin{array}{|ccccc|}
			\hline
			\Delta \phi&E_{trap}&{KE_{trap}}& {PE_{trap}}&{IE_{trap}} \\
			\hline
			0&-84.5&44.1&-78.0&-50.6\\
			0.9\pi&-110.1&57.4&-93.9&-73.6\\
			\hline
		\end{array}
	\end{math}
	\caption{Trapped energy components of the two-soliton ejection stimulated by two in-phase solitons $\Delta \phi=0$ and two solitons with phase difference $\Delta \phi=0.9\,\pi$. Parameters: $u_0=\sqrt{2\,V_0}$, $V_0=16$, $\alpha=\sqrt{2\,V_0}$, $x_0=-3$, $v_i=0.1$,  $g=1$.}
	\label{table3}
\end{table}

\section{Conclusions and outlook}
\label{concsec}
We have shown that soliton ejection may be obtained as an outcome of the bright soliton scattering by a modulated reflectionless potential well. It was found that with a perfect reflectionless potential well, no soliton ejection occurs for the whole range of potential and incident soliton parameters. For soliton ejection to occur, the width of the potential well had to be increased while keeping its depth fixed which represents a modulated potential well that deviates from the reflectionless form. As a result of this deviation, an amount of radiation is always emitted together with soliton ejection.  It was found, though, that minimum radiation is produced when the energy of the incoming soliton is resonant with the energy of trapped modes of the potential well. Within this general setup, detailed investigations have characterised the soliton ejection in terms of all parameters involved such as the incident soliton's initial speed, initial position, and initial amplitude, in addition to the effect of the potential depth. The investigation has identified the potential depth and input soliton amplitude as the most effective parameters that can be used to generate high-speed soliton ejection.

A theoretical model was established in order to understand the physics underlying the ejection mechanism and to predict the ejection speed. The model is based on energy exchange between the incident soliton and a trapped mode. We have used a known localised trapped mode for an integrable version of the NLSE with the P\"oschl-Teller potential. This enabled us to calculate the ejection speed essentially in terms of the depth of the potential well. However, it turned out that although radiation is typically small, it needs to be taken into account in order to have an accurate estimate of the ejection speed. The resonant case of minimum radiation is an exception, where we have seen that radiation has a marginal effect on the calculated ejection speed. While our theory accounts accurately for the trapped mode and the ejected soliton, it does not account for the emitted radiation.  We believe, a future revisit of this problem may result in such an account. Based on this model, the following simple picture for the mechanism is developed. An incident bright soliton with norm that is almost equal to twice the norm of the trapped mode and energy almost equal to the energy of the trapped mode, undergoes a transition from its initial state to that of the trapped mode. While the energy of the incident soliton is almost adequate to form the trapped mode, there will be an excess of norm (matter in case of BEC) which is a left-over from this transition. Due to the focusing nonlinearity, a bright soliton will be formed from the left-over norm together with some radiation. The newly-formed soliton will acquire a translational kinetic energy to balance the negative nonlinear interaction energy used to form it, and hence gets ejected.

To verify this understanding, we have considered more complicated setups where the trapped mode is a multi-node state. Higher ejection speeds were obtained in this case since the magnitude of the binding energy increases with the number of nodes. Another interesting setup that we investigated results in a simultaneous ejection of two solitons from both sides of the potential. The sensitivity of the outcome and ejection speed dependence on the relative phase between the two incident solitons was investigated and suggests a tool for soliton phase interferometry.

In conclusion, our work has identified the necessary conditions to achieve soliton ejection, provided controllability on the ejection speed value in terms of the potential and incident soliton parameters, and unveiled the underlying physics of ejection.

\section*{Acknowledgment}
The authors acknowledge the support of UAE University through grants UAEU-UPAR(1) 2019 and UAEU-UPAR(11)-2019.

\newpage



\end{document}